\documentclass[sigconf,balance=false]{acmart}
\usepackage{popets}

\usepackage{colortbl}
\usepackage{adjustbox}
\usepackage{array}
\usepackage{tabularx}
\usepackage{subcaption}
\usepackage{multirow}
\usepackage{makecell}
\usepackage{csquotes}

\setcopyright{popets}
\copyrightyear{YYYY}

\acmYear{YYYY}
\acmVolume{YYYY}
\acmNumber{X}
\acmDOI{XXXXXXX.XXXXXXX}
\acmISBN{}
\acmConference{Proceedings on Privacy Enhancing Technologies}
\begin{document}

\title[Beyond the Request: Harnessing HTTP Responses]{Beyond the Request: Harnessing HTTP Response Headers 
for Cross-Browser Web Tracker Classification in an Imbalanced Setting}


\author{Wolf Rieder}
\email{w.rieder@tu-berlin.de}
\orcid{0009-0001-4932-9814}
\affiliation{%
  \institution{Technische Universität Berlin}
  \city{Berlin}
  \state{}
  \country{Germany}
}

\author{Philip Raschke}
\email{philip.raschke@tu-berlin.de}
\orcid{0000-0002-6738-7137}
\affiliation{%
  \institution{Technische Universität Berlin}
  \city{Berlin}
  \state{}
  \country{Germany}
}

\author{Thomas Cory}
\email{cory@tu-berlin.de}
\orcid{0000-0002-3452-9944}
\affiliation{%
  \institution{Technische Universität Berlin}
  \city{Berlin}
  \state{}
  \country{Germany}
}


\renewcommand{\shortauthors}{Rieder et al.}

\begin{abstract}
    The World Wide Web's connectivity is greatly attributed to the HTTP protocol, with HTTP messages offering informative header fields that appeal to disciplines like web security and privacy, especially concerning web tracking. Despite existing research employing HTTP request messages to identify web trackers, HTTP response headers are often overlooked. This study endeavors to design effective machine learning classifiers for web tracker detection using binarized HTTP response headers. Data from the Chrome, Firefox, and Brave browsers, obtained through the traffic monitoring browser extension \textit{T.EX}, serves as our dataset. Ten supervised models were trained on Chrome data and tested across all browsers, including a Chrome dataset from a year later. The results demonstrated high accuracy, F1-score, precision, recall, and minimal log-loss error for Chrome and Firefox, but subpar performance on Brave, potentially due to its distinct data distribution and feature set. The research suggests that these classifiers are viable for web tracker detection. However, real-world application testing remains pending, and the distinction between tracker types and broader label sources could be explored in future studies. 
\end{abstract}


\maketitle

\section{Introduction}
\label{sec:intro}


The rapid evolution of the Web has cemented its role as an integral component of modern life. Users access a vast array of information and services through desktop and mobile browsers. However, every interaction, from content consumption to mere site visitation, reveals information about the habits and preferences of users, which in turn has implications for personal privacy and online security.


These privacy concerns are highlighted by the ubiquity of behavioural targeting. A detailed behavioural profile can be constructed by analyzing a user's browsing history, facilitating the delivery of tailored advertisements and content. This profiling is made possible by web tracking that employs sophisticated techniques to track users across the Web. Given the increasing concerns about such practices~\cite{melicher,chanchary}, there is a growing emphasis on research in web privacy.


An established countermeasure against unsolicited HTTP communications is the use of filter lists~\cite{traverso} such as \textit{EasyList}~\footnote{\url{https://easylist.to/easylist/easylist.txt}} (EL) and \textit{EasyPrivacy}~\footnote{\url{https://easylist.to/easylist/easyprivacy.txt}} (EP). However, their dependency on manual curation makes them vulnerable to oversights and rapid obsolescence~\cite{hashmi,alrizah,fouadPixel}. Moreover, some trackers employ strategies like domain masking to evade these lists~\cite{chen}.


In an attempt to mitigate these shortcomings, efforts of past decades have explored the potential of Machine Learning (ML) in this domain. Various classifiers, some achieving accuracy metrics in the 80th percentile~\cite{cozza,iqbalML}, have been developed using diverse data sources such as HTTP headers and JavaScript API calls. 


Meanwhile, the potential of information contained in HTTP responses has been largely overlooked, except for HTTP cookies. The value of this information might be doubted because it becomes available only after web tracking has occurred. However, this limitation primarily affects scenarios where the model's predictions are used to block communications with web trackers in real-time. Alternative approaches are feasible: (i) dynamically generating filter lists based on the model's predictions, or (ii) using it as a supplementary tool for classifiers that depend solely on pre-HTTP request information.




Most previously proposed classifiers have not addressed the issue of deployability, suggesting that the field is still in an exploratory phase. Consequently, it is valuable to examine the suitability of HTTP response information for web tracker classification. Unlike HTTP request information, which is controlled by the client and varies between users, HTTP response information is configured by the server, i.e., the potential web tracker. Notably, a study in the related field of web security underscores the effectiveness of this approach~\cite{mcGahagan}. 



Our key contributions are as follows:
\begin{enumerate}
    \item We conduct a study of HTTP response headers across multiple browsers, focusing on their potential for web tracker detection within the Tranco global top 10K websites. Furthermore, we identify challenges with this type of data for developing ML classifiers. 

    \item We present a semi-automated ML pipeline which facilitates the training of ML models for the binary classification of web trackers based on HTTP response or request headers. 

    \item We train and evaluate ten supervised classifiers. Our findings indicate that when applied to a subset of binarized HTTP headers, selected tree and gradient boosting models can detect web trackers. Our best classifier achieves ROC-AUC, AUPRC, and an F1-score exceeding $0.93$. 

    \item We conduct a comprehensive and multi-faceted evaluation of these classifiers, which includes longitudinal studies to gauge potential concept drift and evaluates the cross-browser performance of classifiers trained solely on traffic data collected within the Chrome browser. We show that Chrome data captures a majority of the structural header characteristics of trackers. Our evaluation incorporates 13 metrics, offering a holistic view of the binary classification challenge. In addition, we use our pipeline to train classifiers using only HTTP request headers to highlight the comparative value of HTTP response headers for web tracker detection. 
    
\end{enumerate}

The remainder of this paper is structured as follows: Section~\ref{sec:related-work} reviews past research in the field of web tracker detection. 
Section~\ref{sec:approach} introduced definitions, our research questions, and the analyzed datasets. Section~\ref{sec:data-analysis} explores HTTP response headers from Chrome, Firefox, and Brave datasets, providing general findings and identifying potential challenges for the ML pipeline. Section~\ref{sec:experimental-setup} describes the design and implementation of the ML pipeline, including selected models, metrics, and baselines. Section~\ref{sec:experiments} presents the evaluation of the trained classifiers. The penultimate Section~\ref{sec:discussion} discusses our results and limitations, and Section~\ref{sec:conclusion} concludes the paper.

\section{Related Work}
\label{sec:related-work}


This section provides an overview of existing research on web tracker detection based on machine learning. To highlight how our proposed classifier differs from prior work, we focus this overview on classifiers that rely solely on HTTP protocol information. For the sake of completeness, however, we summarize other approaches at the end of the section.

\subsection{Classification Based on HTTP Information}
\label{cha2:ml}


Approaches to the detection of web trackers with the support of machine learning date back to the early 2000s. In 2010, Yamada et al. \cite{yamada} were the first to use temporal link analysis to build a graph based on hosts related to HTTP requests from the traffic generated by corporate networks. Their classifier considers temporal information, such as the visit duration at a specific website or host. Yamada et al. consider this a discriminating criterion based on the assumption that users spend less time on websites of web trackers, as trackers generally appear as third parties that are contacted briefly in the context of other websites rather than first parties that users actively visit.  


In 2013, Bau et al. \cite{bau} described and discussed the usefulness of machine learning as an effective counter to web tracking. They outlined problems with manually curated filter lists - problems that still hold true today, such as the reliance on manual curation and the inflexibility of static lists. Beyond building an \textit{Elastic Net} model focused on traces of content extraction in the \textit{Document Object Model} and HTTP headers to show the advantages of automated approaches, Bau et al. laid down requirements for future machine learning models as well as problems facing researchers in this domain.


A year later, Bhagavatula et al. \cite{bhagavatula} tried to identify ad-related URLs, a topic closely connected to the web tracking detection problem. To this end, they crawled the Alexa top 500 websites to compile an extensive dataset with six distinct feature sets based on the URLs themselves as well as additional metadata. Their analysis approach is based predominantly on \textit{Support Vector Machines} (SVMs) with different kernels, achieving average accuracy and precision scores above 89\%. Like Bau et al. before them, Bhagavatula et al. recognized the need for machine learning models to replace filter lists and put forward additional requirements for production-ready machine learning approaches, such as optimized training times.


In 2015, Gugelmann et al. \cite{Gugelmann2015} proposed an alternative approach by deriving several statistical features from HTTP traffic, e.g., up- and download volume. In contrast to prior studies, they thoroughly described their model-building process and emphasized their feature engineering. Their feature selection process included the use of Pearson's correlation coefficient as well as the consideration of a feature's entropy. In addition, they employed an unspecified classification tree, among other classifiers, to find an optimal Naïve Bayes model that achieved a precision of 83\% and was able to correctly identify previously unseen trackers.


In the same year, Li et al. \cite{li} proposed \textit{TrackAdvisor}, a system to classify HTTP requests based on the collective statistics of cookies contained therein. Using \textit{FourthParty}, they captured 563,031 requests and identified 22,270 cookies sent to third parties. Rather than relying solely on filter lists, they used a multi-step process emphasising manual inspection to label the captured data. Notably, the training and test sets created from the captured data comprised just 500 requests each and had a balanced class distribution, which differs from the distribution found in the original data. Nevertheless, Li et al. used these sets to train an SVM that ultimately achieved a precision of 99.4\% and a recall of 100\%. Like Gugelmann et al., they point out the usefulness of the HTTP \textit{Referrer} header for third-party tracker detection.


In 2016, Dudykevych and Nechypor \cite{dudykevych} also used cookies as the basis for their classifiers. In contrast to Li et al., however, they developed their own crawler, used a larger dataset of 550,000 cookies, and developed a different feature set. The latter is only briefly described, but information from the cookies and HTTP requests were used as features. It can be assumed that the features are numerical and categorical, as the authors used Principal Component Analysis and correlation, which require numerical data, as well as binarization of categorical features. The resulting classification accuracies ranged from 82.01\% for Naïve Bayes to 95.43\% for the regularized logistic regression method.


The analysis of HTTP traffic for and with machine learning models is also an important topic in web security research, albeit with a different target (i.e., label) than web privacy research. In a 2019 study, McGahagan et al. \cite{mcGahagan} analyzed response headers from 45\,874 websites and extracted 672 features, resulting in a final feature vector with 22 properties. In addition, they tested whether over- and undersampling affected the model's performance. Their extensive coverage of metrics, models, different training and tuning combinations, and feature engineering documentation can be considered standard among ML practitioners but is somewhat rare among related research in the field of web tracking detection. Crucially, an important contribution of this paper is the in-depth presentation of the usefulness of HTTP response headers for ML models to detect malicious entities. We apply a similar approach but to web privacy research. 


Unlike most papers discussed in this section, which have used data from the Firefox browser to test the performance of supervised learning models, Din et al. \cite{din} used deep learning (DL) techniques for \textit{Brave} and other \textit{Chromium}-based browsers. Although their approach is geared towards the image-based detection of advertising in websites rather than web tracking, their incorporation of multiple browsers is highly relevant for models that classify web trackers.


Cozza et al. \cite{cozza} developed a system that not only classifies HTTP requests but also automatically updates the filter list. Although this study is not the first to detect previously undocumented trackers using machine learning techniques, the deployment of a tracker-detecting browser extension that combines a filter list with a machine learning model and updates its ground truth marks an important step forward in tracker detection technology.


Guarino et al. \cite{guarino} built upon prior research by Gugelmann et al. and Li et al. by training multiple models based on combinations of features used by these studies, including different HTTP headers and the ratio of upstream and downstream data transmissions. Their top performer is a Random Forest model with an average accuracy of 92\%. It should be noted, however, that this study is based on a relatively small dataset of just 1000 observations. Nevertheless, the authors astutely emphasize that content delivery networks (CDNs) are a primary cause of false positives in the domain of web tracker detection due to similarities in traffic patterns between CDNs and third-party trackers.


Taking advantage of the observation that HTTP-based communication between clients and servers often comprises cascades of multiple HTTP messages, Iqbal et al. \cite{khaleesi} conducted an extensive study that uses entire request chains rather than individual requests as the basis for classifying web tracking. Notably, they were among the first to consider responses alongside requests by developing three distinct feature sets: the first is based on the entire request chain, the second relies on information pertaining to individual requests, and the third considers response headers and metadata. By training a Random Forest model on five distinct datasets with different configurations from separate crawls, Iqbal et al. achieved accuracies above 98\%. However, if the model was trained on only one crawl and tested on the others, the accuracy dropped by up to 14\%, highlighting the difficulties of applying models to datasets with varying configurations, e.g., data collected across different browsers. Whereas Iqbal et al. selected features based on intuition and domain knowledge, our approach selects features from a data-driven standpoint and considers all response headers for feature extraction.

\subsection{Other Detection Approaches}


Beyond the HTTP-focused approaches outlined above, various classification methods utilizing different data sources to extract distinctive criteria have been proposed and evaluated. These include classifiers focused on JavaScript code and API access events that aim to identify device fingerprinting techniques, pure URL-based classifiers, CNAME cloaking detectors, and graph-based classifiers that leverage the topology of the Web to identify third-party tracking in particular. The following paragraphs briefly highlight notable research on these alternative approaches.



Focusing on JavaScript API calls, Wu et al.~\cite{dmtTrackerDetector} developed \textit{DMTTrackerDetector}, a system using decision trees in which observed API calls serve as features. 
Their system, comprising a 717-dimen- sional feature vector, achieved 97.8\% accuracy and addressed obfuscation issues, showing the distinct JavaScript API usage by trackers and non-trackers. 


Bahrami et al.~\cite{bahrami} refined the scope of their analysis by focusing on device fingerprinting rather than API calls in general. Their approach combines graph analysis with supervised and unsupervised learning to detect web trackers in historical JavaScript file data collected between 2010 and 2019 by creating a graph that includes temporal aspects and the output of \textit{Abstract Syntax Tree}-based keyword extraction. Their method, which includes nine graph-related features and graph embedding, as well as a clustering model, yielded accuracies up to 89.13\% and identified distinct clusters for fingerprinting and functional APIs, highlighting the viability of such an approach.


Kalavri et al. \cite{kalavri} also proposed a graph-based approach, using neighborhood analysis and label propagation on a bipartite graph representing connections between first-party websites and third-party services to detect web trackers, achieving over 97\% accuracy. They analyzed six months of data, finding that most trackers are closely linked, with precision scores ranging from 64 to 92\%.


Combining these graph-based approaches, Castell-Uroz et al.~\cite{castell} proposed a tripartite network graph to identify third-party tracking resources and their origin domains within first-party websites. By applying this approach to a dataset comprising over 90,000 domains, they detected new trackers through their \textit{hash value popularity} and \textit{dirt level}, finding that higher values correlated with the likelihood of a given resource to be tracking-related.


Conversely, Metwalley et al.~\cite{metwalley} developed an unsupervised algorithm analyzing URL queries and HTTP request headers for identifying trackers. Their method, applied to 200 Alexa-ranked websites, identified 34 new trackers and successfully supported the analysis of cookie-matching practices, though it had limitations in detecting all known trackers.



\subsection{Summary}

Research presented in this section explored various approaches to detect web trackers using machine learning, utilizing various available information. However, HTTP response information has been mostly neglected, and if considered, only the cookie information or a small subset of header information was taken into account. This gap may stem from the intuition that the model's predictions should be utilized in a real-time application scenario, ignoring use cases where ex-post detection of trackers is sufficient. However, prior studies often owe a thorough discussion on the deployability of their approaches. Related to this, training and test data have been generated by the same browser, often in the same browsing session. The generalization of a model's predictions is yet to be evaluated. Furthermore, the metrics used to assess a classifier's performance are not homogenized in the related literature, making a direct comparison of the classifiers' performances unfeasible. 



\section{Approach}\label{sec:approach}

This study addresses the research gap identified in our review of related research by evaluating the (cross-browser) performance of commonly used machine learning classifiers that are trained on HTTP response headers. In this section, we present our underlying research questions and the approach we employ to address them.

\subsection{Preliminaries}
A core principle of automated web tracker detection is delineating what constitutes a \textit{tracker}. The focus of this delineation lies on the tracking entities and not on the tracking activity. Our definition of a tracker (T) and non-tracker (NT) is contingent upon the ground truth used to label each dataset: 

\begin{align*}
H & : \text{Set of all HTTP responses' remote hostnames.} \\
R & : \text{Set of detection rules derived from EL and EP.} \\
\text{match}(h, r) & : 
\begin{cases} 
1 & \text{if } h \text{ matches rule } r \\
0 & \text{otherwise}
\end{cases} \\
T & = \{ h \in H \mid \exists r \in R : \text{match}(h, r) = 1 \} \\
NT & = \{ h \in H \mid \forall r \in R : \text{match}(h, r) = 0 \}
\end{align*}

In other words: if the remote hostname (i.e., the sender) of a response matches at least one rule in our aggregate list compiled from the EL and EP filter lists, we label the response as a \textit{tracker}.

For the sake of simplicity, we use the terms \textit{header} and \textit{response header} interchangeably to refer to HTTP response headers throughout the remainder of the paper unless explicitly stated otherwise.


\subsection{Research Questions}
The primary objective of this study is to assess the efficacy of response headers for automated web tracker detection. Based on the ML approaches of prior research and the observed areas to further extend our empirical understanding and detection of web trackers, we formulate the following research questions (RQ) that guide us throughout this study: 
\begin{itemize}
    \item \textbf{RQ1: What are the key characteristics of HTTP response header data in the web tracking domain?}
    We want to understand the datasets to develop our pipeline and classifiers accordingly.  
    
    \item \textbf{RQ2: Can trained classifiers detect web trackers with only response header features in an imbalanced setting while achieving a similar performance to past research?} 
    We aim to train high-performing classifiers while keeping the likely true distribution. We define our lower bound as an F1-Score of 0.9 based on the classification results of past research to provide a meaningful addition to existing detection solutions. Trackers and non-trackers are not equally present in the datasets, thus an imbalance is present and we want to assess how well classifiers perform in this setting. 
    
    \item \textbf{RQ3: How large is a classifier's performance degradation when it is trained on one browser's HTTP traffic data and then applied to other browsers?}
    Web trackers might exhibit different header characteristics depending on the browser. We use data from three browsers to capture a potential variety of web trackers and to assess the stability and generalization stability of our classifiers. 
\end{itemize}

We address our RQs by developing a semi-automated ML pipeline that transforms HTTP traffic datasets collected with \textit{T.EX - The Transparency EXtension} (T.EX) \cite{philipRaschke} and use it to train several supervised binary classifiers. Additionally, we use reliable data compiled by prior research to facilitate future comparisons between new classifiers. Semi-automated means that the ML pipeline has to be started manually, is based on manual data exploration, and does not employ any \textit{AutoML} libraries. The pipeline executes all steps automatically but cannot react to structurally new datasets without any changes or human intervention. Fully automating the pipeline requires the development of a direct interface between T.EX and the pipeline, i.e., automatically running new crawls and importing the new datasets to the pipeline, as well as structural converters for datasets from other sources.

To test the robustness and reliability of our classifiers, we train them on a Chrome dataset and apply them to other browser datasets (latitudinal comparison) and another Chrome dataset generated one year later (longitudinal comparison). Furthermore, as past research observed an uneven distribution of trackers and non-trackers, we train our classifiers with the original imbalanced distribution to assess how effectively our classifiers perform on the true distribution.  

\newcommand{\ra}[1]{\renewcommand{\arraystretch}{#1}}
\begin{table*}\centering
\small
\ra{1.3}
\begin{tabular}{@{}rrrcrrcrrcrr@{}}\toprule
& \multicolumn{2}{c}{$\text{Chrome}_{22}$} & \phantom{abc}& \multicolumn{2}{c}{$\text{Firefox}_{22}$} &
\phantom{abc} & \multicolumn{2}{c}{$\text{Brave}_{22}$} &\phantom{abc} & \multicolumn{2}{c}{$\text{Chrome}_{23}$}\\ 
\cmidrule{2-3} \cmidrule{5-6} \cmidrule{8-9} \cmidrule{11-12}
& T & NT && T & NT && T & NT && T & NT\\ \midrule
\#Responses & 256,202 & 602,516 && 309,391 & 526,865 && 2969 & 561,694 && 206,747 & 596,066\\
\#Responses in \% & $\approx0.3$ & $\approx0.7$ && $\approx0.37$ & $\approx0.63$ && $\approx0.5$ & $\approx99.5$ && $\approx0.26$ & $\approx0.74$\\
\#Unique headers in total & 2213 & 7577 && 2516 & 7322 && 237 & 7348 && 2164 & 7398\\
Unique Headers per response ($Q1, \tilde{\mu}, Q3$) & 10,13,16 & 12,15,19 && 12,15,18 & 13,16,19 && 13,18,20 & 12,15,19 && 10,13,16 & 13,15,19\\
\bottomrule
\end{tabular}
\caption{Overview of the collected datasets from \cite{zenodo}, split by datasets as well as tracker (T) and non-trackers (NT).}
\label{tab:summary-table}
\end{table*}

\subsection{Data Collection}
We use existing datasets containing labeled HTTP messages by EL and EP. Following best practices suggested by \cite{demir}, the crawls that generate the datasets should fulfill three criteria: (i) repeated crawls across multiple browsers, (ii) documentation of the geographical region as well as the technical environment, and (iii) selection of web sites is based on the Tranco ranking, which was proven to be less biased than the Alexa ranking \cite{pochat}. An open-source dataset by Raschke that used T.EX fulfills all three criteria and is described as follows: 

\begin{quote}\small
Multiple simultaneous and stateful crawls (6 x Chrome, 6 x Brave, 6 x Firefox) of the Tranco top 10K websites (as of 16th of May 2022) were performed on the 12th of August 2022 with T.EX [...]. Measurements had been carried out on 18 Amazon Web Services instances (c5.large) running Windows Server. All instances were launched in Frankfurt, Germany (eu-central-1). 

After completing the crawls, each instance's extension storage was extracted [...]. Note: 2 of the 6 Firefox instances crashed during the crawl. Therefore, only 4 Firefox datasets are available \cite{zenodo}. 
\end{quote}


To assess potential concept drift and resulting model degradation over time, we crawled one more dataset for Chrome in 2023 and made it open-source \cite{zenodo-new}. We follow a similar methodology as described in \cite{raschke2022}, using T.EX and visiting the landing pages with Chrome to explore the top 10K websites from the Tranco ranking as of March 29\textsuperscript{th}, 2023. 

Our reasons for only generating an additional dataset for Chrome are as follows: firstly, Chrome is the most widely used browser with a market share of 66.13\% on desktops, whereas Firefox has 7.1\%~\footnote{\url{https://gs.statcounter.com/browser-market-share/desktop/worldwide}} and Brave has less than 1\%~\footnote{\url{https://kinsta.com/browser-market-share/}}. Based on this distribution, the Chrome dataset serves as the foundation for training to reflect and reach the majority, i.e., using data most users encounter. Secondly, web privacy research often relies on a simulated Firefox browser with Selenium due to the use of \textit{OpenWPM}, leaving space for other browsers of interest. Thirdly, it helps us answer whether a model based on Chrome data can be applied to Firefox or Brave while achieving high-performance metrics.

\section{Initial Response Header Analysis}\label{sec:data-analysis}

To address RQ1 and gain a better understanding of the data underlying our analysis before designing our ML pipeline, we conduct an initial analysis of the data, the results of which we present in this section. 
Our exploration is guided by intuition and domain knowledge. 
We first explore headers across all three browsers to find empirical evidence of their potential relation to trackers.
The goal is to identify the key characteristics of the datasets and not to present a complete analysis. 
We then use these insights to develop our classifiers and evaluate the effectiveness of response headers for web tracker detection. 

\subsection{Descriptive Analysis of Headers}
First, we conduct a descriptive analysis to understand the data structure and distribution of class and feature features within the datasets. Table~\ref{tab:summary-table} summarizes descriptive metrics for each dataset. We begin with the datasets $\text{Chrome}_{22}$, $\text{Chrome}_{23}$, and $\text{Firefox}_{22}$. All three datasets show similar results for the distribution between trackers and non-trackers, as well as the number of unique headers in total and per response. 

Let $B$ represent the set of browsers, where $B = \{ \text{Chrome}_{22}, \allowbreak \text{Chrome}_{23}, \allowbreak\text{Firefox}_{22}, \text{Brave}_{22} \}$ and let $L$ represent the set of labels, where $L = \{T, NT\}$. For each browser $b \in B$ and each label $l \in L$, we define $HN_{b,l}$ as the set of all distinct header names across all HTTP response messages from a browser $b$ and categorized under label $l$. The number of unique headers for each combination of $b$ and $l$ is defined as the cardinality of $HN_{b,l}$: 
\[ |HN_{b,l}| = |\{ hn \mid hn \in HN_{b,l} \}| \]

We can observe a small quantitative change in the number of unique headers between $\text{Chrome}_{22}$ and $\text{Chrome}_{23}$. 
On average, non-trackers have more unique headers as well as a higher median $\tilde{\mu}$ and interquartile range than trackers. 

As Brave is a privacy-oriented browser that performs extensive blocking of web tracking by default, $\text{Brave}_{22}$ exhibits a far lower number of trackers and unique headers in tracking-related response messages than the other datasets. Interestingly, the $\tilde{\mu}$ equals 18 for trackers and 15 for non-trackers, representing the only instance where trackers had a higher number. Although the dataset comprises almost exclusively non-trackers, the number of unique headers is similar to the other datasets. 

\begin{figure}[htbp] 
    \centering
    \includegraphics[width=1\columnwidth]{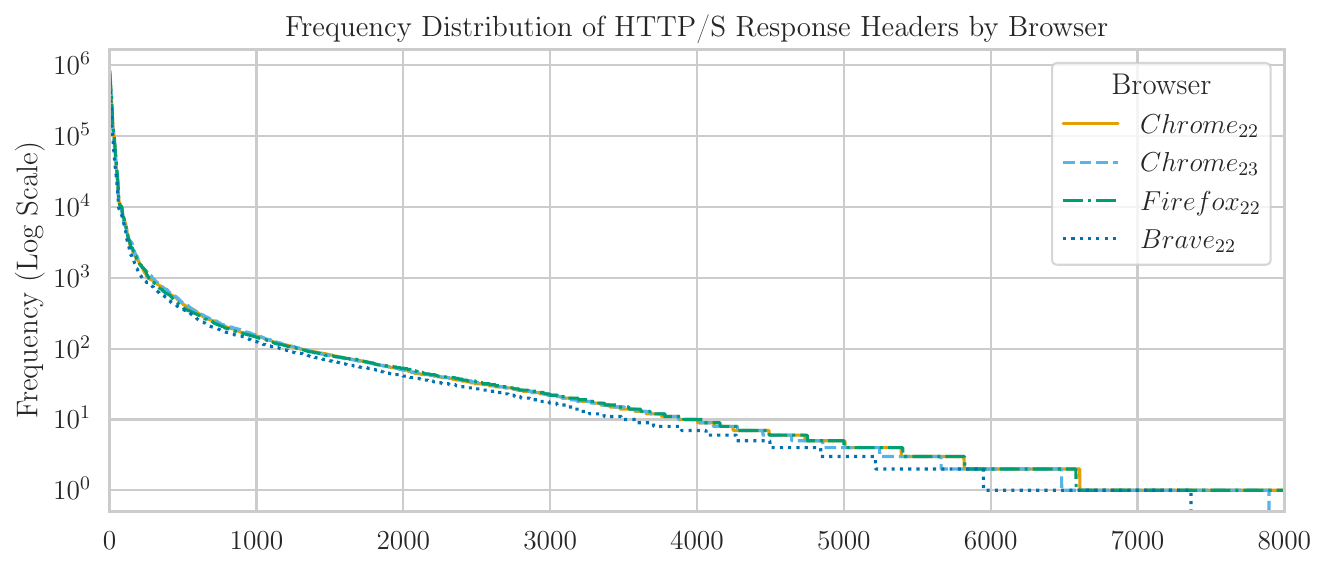}
    \caption{Frequency distribution of headers for each browser, indicating the frequency with which each header is present in the dataset.}
    \label{fig:freq_dist_headers}
    \Description[Frequency distribution of headers for each browser, indicating the frequency with which each header is present in the dataset.]{In this frequency plot we can observe that only a few number of headers are prevalent and most headers are rarely seen in the wild.}
\end{figure}

Figure~\ref{fig:freq_dist_headers} shows a frequency plot to outline how often each header appears in the dataset. The plot demonstrates a power-law distribution, where few headers occur frequently, and many others form a long tail of low frequency. 
This sparseness indicates that while there are many different headers, only a few are consistently used across different websites and by different browsers. 


\begin{figure}
    \centering
    \includegraphics[width=0.75\columnwidth]{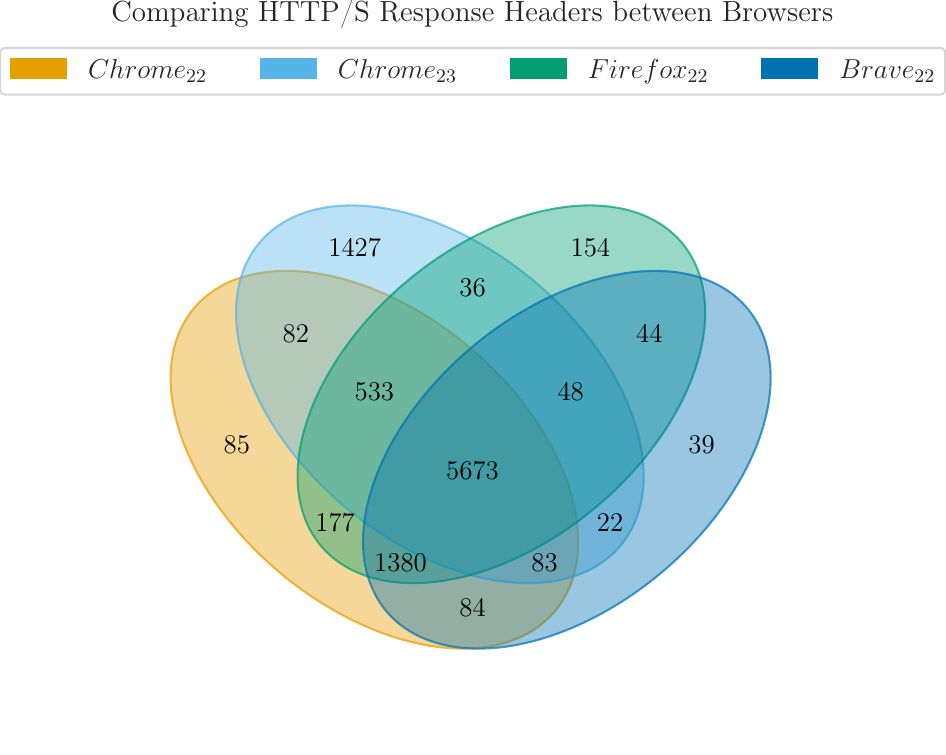}
    \caption{Venn diagram of HTTP response headers in the examined datasets, revealing a core set of 5672 headers that are common across Chrome, Firefox, and Brave. Unique headers are present in each browser, although $\text{Chrome}_{23}$ has the highest number of unique headers, which may reflect browser-specific features or experimental headers.}
    \label{fig:venn_dig}
    \Description[Venn diagram of HTTP response headers.]{Venn diagram of HTTP response headers in the examined datasets, revealing a core set of 5672 headers that are common across Chrome, Firefox, and Brave. Unique headers are present in each browser, although $\text{Chrome}_{23}$ has the highest number of unique headers, which may reflect browser-specific features or experimental headers.}
\end{figure}

Figure~\ref{fig:venn_dig} depicts a Venn diagram that compares the response headers present in the examined datasets, highlighting data and feature similarity between the different browsers. The intersection numbers, such as the 5672 headers common to all browsers, suggest a significant overlap in features across different platforms. However, the presence of headers unique to each browser or shared by only two browsers
indicates that while there is a core set of shared features, there is also a notable amount of variation that may be critical for web tracker detection. 

\begin{figure}
    \centering
    \includegraphics[width=1\columnwidth]{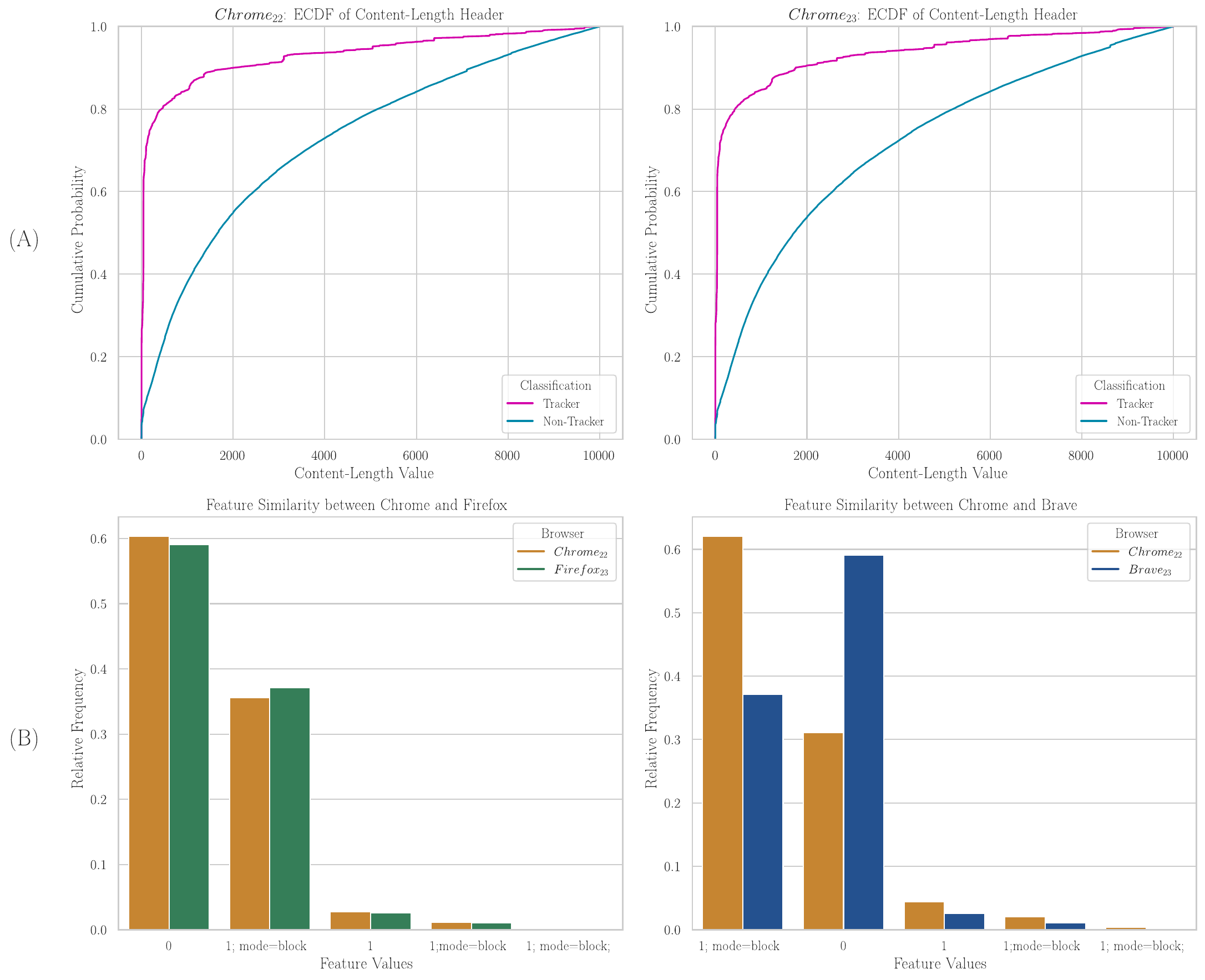}
    \caption{Row (A) presents the ECDFs for both Chrome datasets for the \textit{Content-Length} header values. We set a cut-off value at 10,000 to highlight the core observation and more than half of the responses have a value below this threshold -- $\tilde{\mu}_{T22} = 62$, $\tilde{\mu}_{NT22} = 8068$ and $\tilde{\mu}_{T23} = 45$, $\tilde{\mu}_{NT23} = 8021.5$. Row (B) shows how similar the values across $\text{Chrome}_{22}$, $\text{Firefox}_{22}$, and $\text{Brave}_{22}$ are for the \textit{X-XSS-Protection} header.} 
    \label{fig:cl_header}
    \Description[ECDF plots for two selected response headers.]{Row (A) presents the ECDFs for both Chrome datasets for the \textit{Content-Length} header values. We set a cut-off value at 10,000 to highlight the core observation and more than half of the responses have a value below this threshold -- $\tilde{\mu}_{T22} = 62$, $\tilde{\mu}_{NT22} = 8068$ and $\tilde{\mu}_{T23} = 45$, $\tilde{\mu}_{NT23} = 8021.5$. Row (B) shows how similar the values across $\text{Chrome}_{22}$, $\text{Firefox}_{22}$, and $\text{Brave}_{22}$ are for the \textit{X-XSS-Protection} header.}
\end{figure}

For instance, the \textit{Empirical Cumulative Distribution Function} plots for $\text{Chrome}_{22}$ and $\text{Chrome}_{23}$ presented in Figure~\ref{fig:cl_header} row (A) show that the \textit{Content-Length} values of non-trackers are generally higher than those of trackers, which suggests that non-trackers tend to send larger responses. This is consistent across both datasets, indicating that while browsers evolve, certain characteristics, like the smaller size of tracker responses, remain consistent. Conversely, row (B) of Figure~\ref{fig:cl_header} plots the feature similarity for the \textit{X-XSS-Protection} header, showing that header values can be very similar between the datasets, i.e., the training and test sets. This further supports our observation that most of the examined datasets are similar and that trackers and non-trackers differ in their use of HTTP header fields and values. Although we have presented only a few selected examples here, they highlight the general challenges that come with response message datasets. 

Additional characteristics of the datasets include a high amount of missing header values, resulting from a large number of available HTTP headers, of which any given response only uses a small number. This is exacerbated by the common practice of using custom headers that are not part of the HTTP standard, i.e., are not registered in RFC 4229\footnote{\url{https://www.ietf.org/rfc/rfc4229.txt}}. This results in the presence of 8058 custom headers in the $\text{Chrome}_{22}$ dataset, contrasted by a comparatively small $\tilde{\mu}$ of unique headers per response as shown in Table~\ref{tab:summary-table}. 



The examined datasets are formatted as a matrix, with rows representing individual HTTP requests or responses and each unique HTTP header denoted in a separate column. This, combined with the observed large total number of unique headers and small number of headers per response, results in a sparse matrix where most elements are empty.



\section{Methodology}\label{sec:experimental-setup}

Based on the findings of our data exploration presented in Section~\ref{sec:data-analysis}, we propose a comprehensive methodology to assess the effectiveness of classifiers based on response headers. 

\subsection{Problem Formulation}
We consider the problem of binary classification, where we aim to predict the presence of web trackers based on response headers. The headers, represented by the feature vector $x \in \mathbb{R}^d$, are nominal and can be discrete and continuous variables.
These variables are derived from HTTP headers of four different browser types, denoted as the previously defined set $B$. 
For this study, we specifically consider the datasets Chrome$_{22}$, Chrome$_{23}$, Firefox$_{22}$, and Brave$_{22}$. 

Given a training set $\mathcal{D}_{\text{train}} = \{(x_1 y_1),\ldots,(x_n,y_n)\}$, where $x_i$ is a feature vector of the $i$-th HTTP header and $y_i \in \{0, 1\}$ represents the associated class label, our goal is to train a set of classifiers $f_j: \mathbb{R}^d \rightarrow \{0, 1\}$ for $j \in M$ that minimizes the empirical risk: 

\begin{equation}
R(f_i) = \frac{1}{N}\sum_{i=1}^{N} L(y_i, f_i(x_i)), \quad j \in \{1, \ldots, M\},
\end{equation}

where $L(y_i, f_i(x_i))$ is the loss incurred by predicting $f_i(x_i)$ when the true label is $y_i$. The specific form of the loss function $L(\cdot)$ will be dependent on the classifier $f_j$ used. 

We consider datasets represented by $D_s$, where $s$ can pertain to training, testing, or calibration subsets of the data. We observe several challenges (CH) and key characteristics from our data exploration, which answer our RQ1: 
\begin{itemize}
    \item \textbf{(CH1) High Cardinality:} The feature vectors $x_i$ in our datasets $D_s$ may exhibit high cardinality for certain features. This problem may pose challenges in terms of model training and encoding. 

    \item \textbf{(CH2) Sparseness:} As discussed above, the feature vectors $x_i$ in our datasets $D_s$ are represented as sparse vectors, with each vector defined as:
    \[x_i = \{v_1, v_2, \ldots, v_d\}\]
    A significant proportion of the entries \( v_j \) within \( x_i \) are either zero or represent missing data, resulting in an overall sparseness of $D_s$. Such sparseness may introduce computational complexities and challenge the performance of certain ML algorithms that are not optimized for sparse data. 

    
    \item \textbf{(CH3) Class Imbalance:} For datasets collected from a given browser $B$ and represented as $D_s$, the class distribution may be imbalanced. The majority class is denoted by $\alpha_B$, and the minority class is represented by $\beta_B$, satisfying $\alpha_B + \beta_B = 1$. The degree of this imbalance may vary based on the browser, as shown in Table~\ref{tab:summary-table}, with some datasets exhibiting a more pronounced disparity between $\alpha_B \text{ and } \beta_B$. Such imbalance can introduce bias to models, favoring the majority class and potentially undermining the accurate prediction of the minority class. 

    
    \item \textbf{(CH4) High Dimensionality:} With a high dimensionality $d$ and a high number of observations $n$ per dataset, the classification is susceptible to the \textit{Curse of Dimensionality}, potentially leading to overfitting and necessitating intricate dimensionality reduction techniques. Therefore, applying a process that achieves a reduced feature space $X_i' \subseteq X$ is necessary, entailing a reasonable training time. 
    
    \item \textbf{(CH5) Data Similarity:} The datasets collected across brow- sers can exhibit pronounced similarities in their feature distributions and values. 
    Given the high dimensionality of our feature space, exhaustive pairwise comparisons can be computationally expensive and time-consuming, highlighting the challenge of accurately gauging data similarities in such a complex space. 
\end{itemize}

\begin{figure*}[htbp]
    \centering
    \includegraphics[width=1\textwidth]{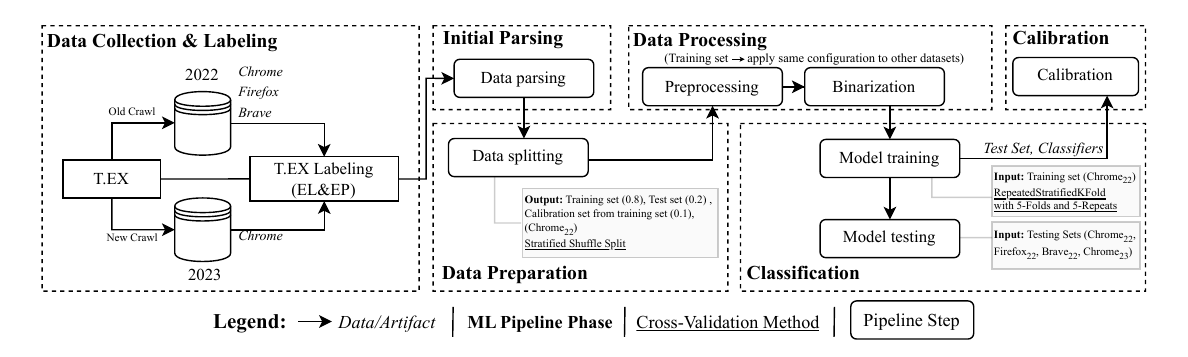}
    \caption{The parameterized pipeline automatically processes the collected datasets and trains the classifiers.}  
    \label{fig:ml-pipeline}
    \Description[The parameterized pipeline automatically processes the collected datasets and trains the classifiers.]{The parameterized pipeline automatically processes the collected datasets and trains the classifiers. The pipeline starts with data collection and labeling. After the data is initially parsed, it is prepared and processed. Lastly, the models are trained and tested, and calibrated.}
\end{figure*}

\subsection{Machine Learning Pipeline}\label{ml-pipeline}

This subsection outlines our ML pipeline that addresses the challenges in classifying web trackers. Related work uses various approaches to detect web trackers, a task closely related to web security research. Both research fields exploit HTTP data, typically requiring extensive data processing and feature engineering to identify useful features. However, Laughter et al. \cite{laughter} explored whether binary features that represent the mere presence of HTTP headers could effectively detect malicious HTTP requests. Their classifiers achieved accuracy scores above 90\%. Similar findings were reported by \cite{bhagavatula} and \cite{dudykevych}. 

The binarization of HTTP headers is appealing for two reasons: (i) it simplifies feature engineering, which is complex for categorical header data, and (ii) it enhances model deployment for tracker detection, as checking header presence is faster than detailed processing. Therefore, we apply and evaluate this method in our analysis. 


We train our classifiers on the $\text{Chrome}_{22}$ dataset, using other datasets for testing as per RQ3. Therefore, only the training set is used to filter out headers, and the results of all steps in the data processing phase in Figure~\ref{fig:ml-pipeline} are afterwards applied to all testing sets. Each step in our pipeline aims to reduce dataset dimensions, addressing overfitting and sparsity issues. 

To handle sparsity and high dimensionality and simplify the data matrix, we use several methods: (i) applying a low variance filter to remove headers with a single value, (ii) eliminating rare headers with a missing-value ratio approaching one, (iii) dropping headers associated with only one label as they might be correlated to the label, and (iv) employing \textit{Approximate string matching} (\textit{Fuzzy Matching}) to merge similar headers based on a weighted ratio of distance metrics (e.g., \textit{Damerau Levenshtein} and \textit{Hamming}).

Fuzzy matching is especially useful due to the presence of hundreds of similar headers within our data. Our initial data analysis discussed in Section~\ref{sec:data-analysis} identifies many examples where headers with similar purpose were written or formatted differently, e.g., \textit{content-lenght}, \textit{cteonnt-length}, or \textit{ntcoent-length}, instead of \textit{content-length}. Therefore, comparing the header names is insufficient, so we automatically assess the value similarity to reduce potential mismatches.

Although many different resampling techniques exist to combat class imbalance, such as over- and undersampling \cite{sampling}, generative adversarial networks \cite{gan}, or variational autoencoders \cite{autoencoders}, the question of whether this step is necessary remains. 
Some disadvantages of such methods are, for example, the removal of observations, the replication of existing observations, or the creation of synthetic samples. 
Selecting and applying a sampling method can lead to overfitting and over-optimism as well \cite{8492368}, notwithstanding the question of whether it is even feasible to train the models on a distribution that was not observed in real-world data. 

Addressing CH3, our pipeline preserves the observed distribution as the imbalance is mild~\footnote{\url{https://developers.google.com/machine-learning/data-prep/construct/sampling-splitting/imbalanced-data}} in the $\text{Chrome}_{22}$ dataset, i.e., a sufficient number of samples exists for each class, which allows proper training to differentiate between classes. 
We also compute metrics that account for or are unaffected by imbalance, and the selected ensemble models are inherently robust to this issue. 



Last, following the findings of Niculescu-Mizil and Caruana~\cite{calibration}, we apply \textit{Isotonic Regression} instead of \textit{Platt Scaling} to calibrate the probabilities calculated by our models. This is done by fitting an Isotonic Regression model to the models' predictions as well as the true labels to adjust the predicted probabilities. The calibration function for each model $i$ is denoted as $C_i(f_i(x))$. 

We summarize our pipeline in Figure~\ref{fig:ml-pipeline}, including the data collection. Note that we did not apply any data imputation to keep the final datasets as close to the original collected datasets as possible. In addition, due to the application of binarization as a final step, we do not need to impute values for missing data as we are solely interested in the presence of headers and not their values. 

We do not generate a validation set for hyperparameter tuning because we want to assess the effect of calibration in isolation. Moreover, tuning requires more computational power. However, this might lead to suboptimal configurations of our classifiers, thus a reduced performance.  

\subsection{Classification Models}\label{models}
As discussed in Section~\ref{cha2:ml}, related studies~\cite{cozza, li, bhagavatula, Gugelmann2015} often used similar classifiers despite the broad range of available models. We build upon these established selections, refining our choice of models based on three conditions: (i) the models should be trainable in a reasonable amount of time, (ii) they should be applicable to this specific use case, and (iii) they should preferably be ensemble methods, although we include a handful of non-ensemble models that were used in past research to allow for direct comparisons between our results and those of related studies. We prefer ensemble methods, as they usually perform better than their counterparts \cite{ensemble}, leading to their popularity over the past few years. Nevertheless, to the best of our knowledge, they have not been widely used in past web privacy research, with the exception of Random Forests. In Table~\ref{table:model-overview}, we provide an overview of our selected models and their use in related work. 

\begin{table}[h]
	\centering
        \scriptsize
	\begin{tabular}{@{}rll@{}}
            \toprule
		\thead{\textbf{Model}} & \thead{\textbf{Ensemble}} & \thead{\textbf{Past Research}} \\
		\midrule
		Decision Tree & No & \cite{Gugelmann2015, dudykevych, t.ex-graph}\\ 
            Random Forest & Yes & \cite{cozza, dmtTrackerDetector, guarino, khaleesi, t.ex-graph}, \cite{dudykevych} (U), \underline{\cite{mcGahagan}}\\
            Extra Trees Classifier & No & Not used \\
            Logistic Regression & No & \cite{bhagavatula, Gugelmann2015, dudykevych, dmtTrackerDetector, t.ex-graph}, \underline{\cite{mcGahagan}}\\
            Naïve Bayes (Gaussian or Bernoulli) & No & \cite{yamada, bhagavatula, Gugelmann2015, dudykevych, dmtTrackerDetector, guarino}\\
            Gradient Boosting & Yes & \underline{\cite{mcGahagan}}, \cite{t.ex-graph}\\
            Light-Gradient Boosting Machine & Yes & Not used\\
            Adaptive Boosting & Yes & \cite{t.ex-graph}\\
            Hist-Gradient Boosting & Yes & Not used\\
            eXtreme Gradient Boosting & Yes & \cite{t.ex-graph} \\
		\bottomrule
	\end{tabular}
	\\[6pt]	
 \caption{Overview of the selected ML models and their use in related research discussed in Section~\ref{cha2:ml}. One paper did not specify exactly which ensemble model was used. The most likely model is marked as \textit{Unspecified} (U). Papers from the field of web security are underlined.}
 \label{table:model-overview}
\end{table}

Other supervised learning methods, such as \textit{Multi-layer Perceptrons} \cite{cozza, guarino}, KNN, and especially SVM, could have been used. The latter was successfully applied in multiple papers \cite{cozza, li, bhagavatula, guarino}. However, both suffer in terms of time and computational complexity. SVM, for instance, has a training time complexity of $O(n^2)$, where $n$ is the number of observations \cite{Steinwart2008SupportVM}. In addition, deep-learning models are not considered, as our research focuses on simple, lightweight models. We also do not pursue a graph-based approach as used by several papers \cite{kalavri, castell} because we do not examine connections between responses, e.g., through grouping based on their URL or building a graph based on their order. 



\subsection{Baseline Models}\label{baseline-models}

We compare our classifiers against three baseline models and multiple metrics that we define in Section~\ref{metrics}. We select one tree classifier, the Decision Tree (DT), which is the foundation of six selected classification models. Our other two baselines are Gaussian Naïve Bayes (GNB) and Logistic Regression (LR), two simplistic and relatively robust models. 

It is good practice to compare against a baseline approach that differs from the presented classifiers while using the same dataset. Raschke et al. \cite{t.ex-graph} presented a graph-based approach named \textit{T.EX-Graph} where edge and node attributes and centrality metrics served as features. Furthermore, the authors included information from the responses but did not pursue classifiers based solely on response headers. This comparability should help assess the performance gains and the relevance of response header values for automated web tracker detection, although the authors applied \textit{Synthetic Minority Oversampling} to balance their dataset. We focus on their results stemming from the \textit{FQDN} dataset, where hosts are modeled as fully qualified domain names (FQDN) because the tracker distribution is similar to that of our datasets with a non-tracker to tracker ratio of 68:32. To verify the results and calculate additional metrics not used in the original study, we replicated their experiment.

\subsection{Models using HTTP Request Headers}
Lastly, we compare our proposed approach to HTTP request header-only models. Our reasoning for this comparison is twofold: (i) showing which additional information can be gained from HTTP response headers, thus the usefulness of our approach and (ii) whether the applied technique of binarization is helpful for models using HTTP request headers. 
To enable a direct comparison between request- and response-based classifiers, we apply the pipeline presented in Section~\ref{ml-pipeline} to the request headers present in our datasets and compare the results for each request-response pair. 

\subsection{Evaluation Metrics}\label{metrics}
We measure the performance of our classifiers according to nine metrics. Several of these metrics are common in the web tracking literature, while others are selected due to their applicability to our classification problem and to understand the relationship with the underlying data in more detail. We report on additional metrics, such as the confusion matrix, in the appendix. 
Moreover, we calculate confidence intervals (CI) for these metrics using bootstrap sampling with 599 samples \cite{wilcox2010fundamentals}. 

\begin{table}[h]
	\centering
        \small
	\begin{tabular}{@{}rl@{}}
            \toprule
		\thead{\textbf{Metric}} & \thead{\textbf{Past Research}} \\
		\midrule
		(Avg.) Accuracy & \cite{Gugelmann2015, dudykevych, t.ex-graph, cozza, guarino, khaleesi, bhagavatula}, \underline{\cite{mcGahagan}}\\ 
            Precision & \cite{Gugelmann2015, t.ex-graph, khaleesi, bhagavatula, li}\\
            Recall & \cite{Gugelmann2015, t.ex-graph, khaleesi, li}\\
            F1-Score & \cite{Gugelmann2015}, \cite{t.ex-graph}\\
            AUC-ROC & \cite{Gugelmann2015}, \cite{bhagavatula}, \underline{\cite{mcGahagan}}\\
            Confusion Matrix & \cite{guarino}\\
            MCC & \underline{\cite{mcGahagan}}\\
            AUPRC & Not used\\
            Log-Loss & Not used\\
		\bottomrule
	\end{tabular}
	\\[6pt]	
 \caption{Overview of applied performance metrics and their use in related research discussed in Section~\ref{cha2:ml}.}
 \label{table:metrics-overview}
\end{table}

The first metrics that are commonly used in web privacy research are Accuracy, Precision, Recall, False Positives (FP), True Positives (TP), False Negatives (FN), True Negatives (TN), and F1-Score. 
Two accuracy scores can be calculated: one from the perspective of trackers, the positive label, and one from non-trackers, the negative label. However, accuracy is not useful with imbalanced datasets because a model could always predict the majority class and achieve accuracy with the same value as the distribution of said class. For this reason, we calculate the following additional metrics to quantify the performance of a classifier with an imbalanced dataset: (i) \textit{Balanced Accuracy} (BACC), (ii) \textit{Log-loss Score}, (iii) \textit{Matthews-Correlation-Coefficient} (MCC), and (iv) \textit{Area Under the Precision-Recall Curve} (AUPRC). A further explanation for each metric is reported in the Appendix~\ref{appendix:metrics}.

\section{Experiments}\label{sec:experiments}

\begin{figure*}[htbp] 
    \centering
    \includegraphics[width=0.8\textwidth]{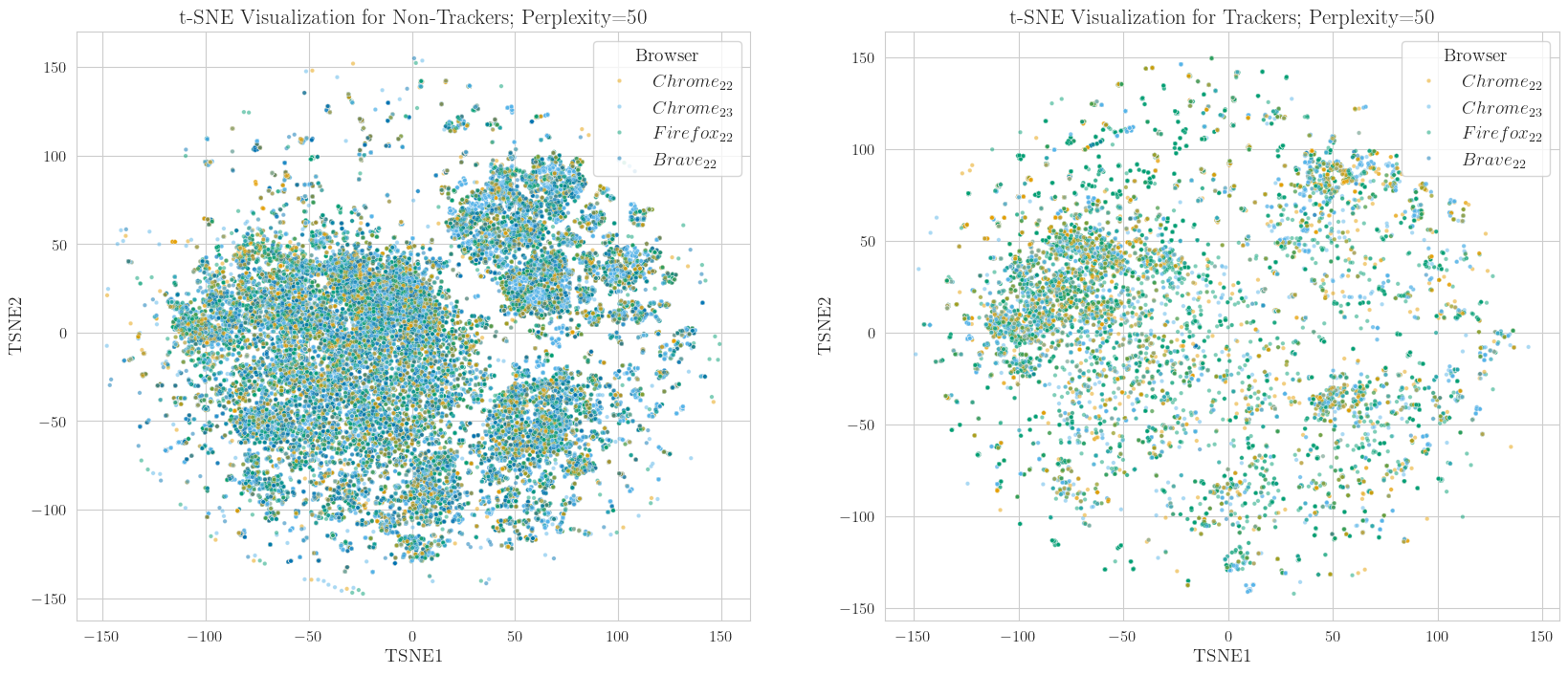}
    \caption{t-SNE plots using a representative sample across all browsers to gauge data similarity. Separate clusters exist for trackers and non-trackers, but there is significant overlap between both classes. The results are separated by class to highlight differences between trackers and non-trackers.} 
    \label{fig:tsne-clustering}
    \Description[t-SNE plots across browsers to gauge data similarity.]{t-SNE plots using a representative sample across all browsers to gauge data similarity. Separate clusters exist for trackers and non-trackers, but there is significant overlap between both classes. The results are separated by class to highlight differences between trackers and non-trackers.}
\end{figure*}

In this section, we evaluate the effectiveness of our proposed metho- dology in detecting trackers. 
We begin by presenting our classifier results based on $\text{Chrome}_{22}$ data and follow up by testing them with data from $\text{Firefox}_{22}$ and $\text{Brave}_{22}$. 
Furthermore, we assess the longitudinal performance with the $\text{Chrome}_{23}$ dataset that was collected almost a year after $\text{Chrome}_{22}$. 
Note that this section only summarizes our results. Performance metrics across all classifiers are reported in the Appendix~\ref{appendix:response}.  

We test our approach on a MacBook Pro (14-inch, 2021) with a 10-core CPU, a 16-core GPU, 1TB of SSD storage, and 32GB of RAM. 

We address CH5 by applying t-SNE to our datasets as shown in Figure~\ref{fig:tsne-clustering}. 
Non-trackers form high-density clusters, indicating that they share similar headers. 
$\text{Brave}_{22}$ exhibits more smaller groups at the edges that are separated from the others. 
Trackers only form one large cluster, showing that commonalities exist even though they are more dispersed overall. This is potentially due to the higher number of custom headers present in tracker responses. Additionally, overlaying the t-SNE plots for trackers and non-trackers reveals a substantial overlap between the two classes.


\definecolor{customblue}{RGB}{112, 189, 255}
\definecolor{customorange}{RGB}{255, 178,112}
\definecolor{customgreen}{RGB}{153,216, 199}

\begin{table*}[t!]
	\centering
        \footnotesize
	\begin{tabular}{@{}rlllllllll@{}}
        \toprule
		\thead{\textbf{Model}} & \thead{\textbf{Accuracy}} & \thead{\textbf{Log-Loss}} & \thead{\textbf{ROC-AUC}} & \thead{\textbf{AUPRC}} & \thead{\textbf{BACC}} & \thead{\textbf{F1-Score}} & \thead{\textbf{Precision}} & \thead{\textbf{Recall}} & \thead{\textbf{MCC}} \\
		\midrule
            \cellcolor{customorange}\textit{LR} & 0.874&0.321&0.829&0.858&0.829&0.773&0.836&0.718&0.69 \\ 
            & [0.872;0.875]&[0.318;0.324]&[0.827;0.831]&[0.856;0.862]&[0.827;0.831]&[0.769;0.776]&[0.831;0.839]&[0.714;0.723]&[0.686;0.693] \\ 
            \cellcolor{customorange}\textit{GNB} & 0.835&3.232&0.803&0.688&0.803&0.723&0.723&0.723&0.605\\ 
            & [0.832;0.836]&[3.193;3.28]&[0.8;0.805]&[0.683;0.693]&[0.8;0.805]&[0.72;0.727]&[0.719;0.727]&[0.719;0.727]&[0.6;0.61] \\ 
            \cellcolor{customorange}\textit{DT} & 0.958&0.324&\cellcolor{customgreen}\underline{0.945}&0.965&\cellcolor{customgreen}\underline{0.945}&0.929&0.945&\cellcolor{customblue}\textbf{0.913}&0.9 \\ 
            & [0.957;0.959]&[0.311;0.34]&[0.944;0.946]&[0.963;0.966]&[0.944;0.946]&[0.928;0.93]&[0.943;0.947]&[0.911;0.915]&[0.898;0.902] \\ 
            RF & \cellcolor{customgreen}\underline{0.959}&\cellcolor{customblue}\textbf{0.162}&\cellcolor{customblue}\textbf{0.946}&\cellcolor{customblue}\textbf{0.98}&\cellcolor{customblue}\textbf{0.946}&\cellcolor{customgreen}\underline{0.93}&\cellcolor{customgreen}\underline{0.947}&\cellcolor{customblue}\textbf{0.913}&\cellcolor{customgreen}\underline{0.901} \\ 
            & [0.958;0.96]&[0.156;0.17]&[0.944;0.947]&[0.979;0.98]&[0.944;0.947]&[0.928;0.931]&[0.949;0.949]&[0.911;0.915]&[0.899;0.903] \\ 
            ET & \cellcolor{customblue}\textbf{0.96}&\cellcolor{customgreen}\underline{0.195}&\cellcolor{customblue}\textbf{0.946}&\cellcolor{customgreen}\underline{0.977}&\cellcolor{customblue}\textbf{0.946}&\cellcolor{customblue}\textbf{0.931}&\cellcolor{customblue}\textbf{0.951}&\cellcolor{customgreen}\underline{0.912}&\cellcolor{customblue}\textbf{0.903} \\ 
            & [0.959;0.961]&[0.186;0.204]&[0.945;0.947]&[0.976;0.978]&[0.945;0.947]&[0.93;0.933]&[0.949;0.953]&[0.91;0.914]&[0.901;0.906] \\ 
            AdaBoost & 0.858&0.673&0.809&0.846&0.809&0.742&0.809&0.686&0.65 \\ 
            & [0.856;0.859]&[0.673;0.673]&[0.806;0.811]&[0.844;0.849]&[0.806;0.811]&[0.739;0.746]&[0.805;0.813]&[0.682;0.69]&[0.646;0.653] \\ 
            GBM & 0.892&0.292&0.85&0.89&0.85&0.804&0.872&0.746&0.734 \\ 
            & [0.89;0.893]&[0.289;0.294]&[0.848;0.852]&[0.888;0.892]&[0.848;0.852]&[0.801;0.808]&[0.869;0.875]&[0.742;0.751]&[0.731;0.738] \\ 
            LGBM & 0.917&0.212&0.881&0.94&0.881&0.851&0.916&0.794&0.797 \\ 
            & [0.915;0.918]&[0.21;0.214]&[0.879;0.883]&[0.938;0.941]&[0.879;0.883]&[0.848;0.853]&[0.913;0.918]&[0.79;0.798]&[0.793;0.801] \\ 
            HistGB & 0.918&0.211&0.884&0.94&0.884&0.854&0.916&0.8&0.801 \\ 
            & [0.917;0.92]&[0.209;0.213]&[0.882;0.886]&[0.939;0.942]&[0.882;0.886]&[0.851;0.857]&[0.913;0.918]&[0.796;0.803]&[0.798;0.804] \\ 
            XGBoost & 0.933&0.182&0.908&0.953&0.908&0.883&0.924&0.845&0.838 \\ 
            & [0.932;0.934]&[0.18;0.184]&[0.906;0.909]&[0.952;0.954]&[0.906;0.909]&[0.881;0.885]&[0.992;0.926]&[0.842;0.848]&[0.835;0.84] \\
            \midrule
        \cellcolor{customorange}\textit{XGBoost} & 0.883 & 2.666 & 0.871 & 0.248 & 0.871 & 0.869& 0.867 & 0.871 & 0.737 \\
        \cellcolor{customorange}\textit{RF} & 0.877 & 0.744 & 0.867 & 0.273 & 0.867 & 0.863&0.860 & 0.867 & 0.726  \\
        \cellcolor{customorange}\textit{GBM} & 0.857 & 1.476 & 0.854 & 0.42 & 0.854 &0.843&0.836 & 0.854 & 0.689  \\
        \cellcolor{customorange}\textit{DT} & 0.843 & 21.54 & 0.827 & 0.332 & 0.827 & 0.824&0.822 & 0.827 & 0.649  \\
        \cellcolor{customorange}\textit{KNN} &0.841 & 1.674 & 0.836 & 0.577 & 0.836 & 0.826&0.819 & 0.836 & 0.655  \\
        \cellcolor{customorange}\textit{AdaBoost} & 0.839 & 0.709 & 0.833 & 0.493 & 0.833 &0.823 &0.817 & 0.833 & 0.649 \\
        \cellcolor{customorange}\textit{SVC} & 0.837 & 0.641 & 0.849 & 0.332 & 0.849 & 0.826 &0.818 & 0.849 & 0.666  \\
        \cellcolor{customorange}\textit{LR} & 0.802 & 12.24 & 0.817 & 0.362 & 0.817 & 0.791&0.785 & 0.817 & 0.601 \\
		\bottomrule
	\end{tabular}
	\\[6pt]	
 \caption{Performance comparison of the $\text{Chrome}_{\textbf{22}}$-trained classifiers for identifying trackers across a subset of 13 metrics. The best results are highlighted in \textbf{bold} and blue, the second best are \underline{underlined} and green, and baseline models are denoted as \textit{italic} and orange. For comparison, the performance results of the \textit{T.EX-Graph} classifier~\cite{t.ex-graph} are reported in the bottom half of the table.} 
 \label{tab:chrome22_results}
\end{table*}

\subsection{Intra-browser Performance}
\begin{figure*}[htbp] 
    \centering
    \includegraphics[width=1\textwidth]{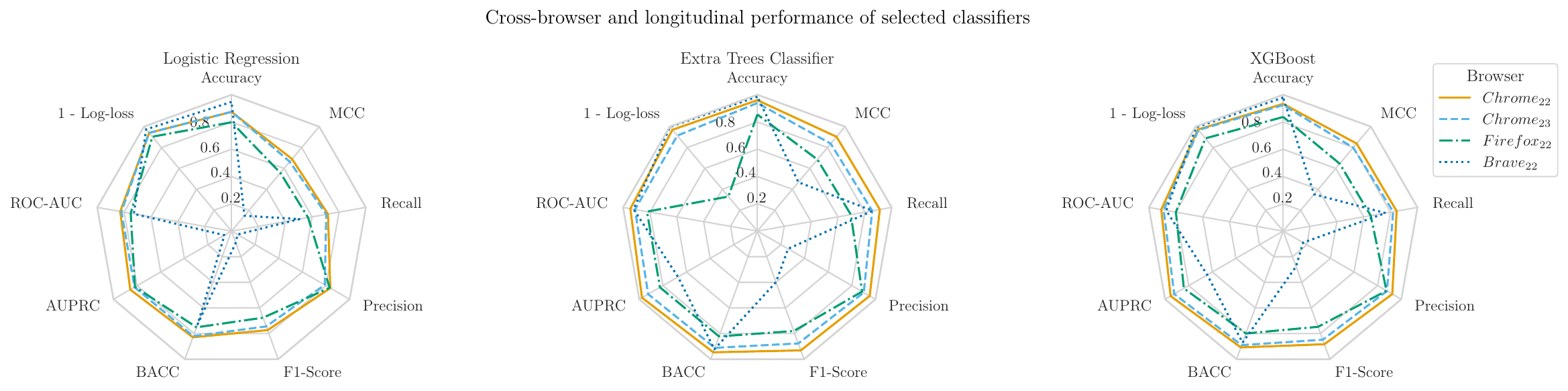}
    \caption{Radar charts plotting the performance of three selected classifiers across a subset of 13 metrics, illustrating that, while all models perform well on the in-distribution test set, the ET model exhibits more consistent performance across the various test sets, suggesting better generalizability. The models show a reduced performance on out-of-distribution test sets, indicating potential overfitting to the training data.} 
    \label{fig:spiderplot}
    \Description[Radar charts of the LR, ET, and XGBoost classifier for cross-browser performance.]{Radar charts plotting the performance of three selected classifiers across a subset of 13 metrics, illustrating that, while all models perform well on the in-distribution test set, the ET model exhibits more consistent performance across the various test sets, suggesting better generalizability. The models show a reduced performance on out-of-distribution test sets, indicating potential overfitting to the training data.}
\end{figure*}

Table~\ref{tab:chrome22_results} shows the performance of our $\text{Chrome}_{22}$-trained classifiers and the baselines. Regarding AUPRC, the RF leads with a score of 0.98, closely followed by Extra Trees (ET) classifier and DT with scores of 0.977 and 0.965, respectively. 

Except for the AdaBoost and GNB, all classifiers show a lower than baseline log-loss value, indicating a high prediction accuracy. The F1-score metric further strengthens the position of ET and RF, as they all score above 0.93, indicating that they are good at differentiating between the positive and negative classes (RQ2). The high MCC scores corroborate the other metrics, placing ET, RF, and DT as the most robust performers with accurate predictions across all categories in the confusion matrix. In contrast, baseline models like LR and AdaBoost perform comparatively lower across all metrics. 

The gradient boosting models, specifically XGBoost, have fairly high scores, as reflected in an AUPRC score of 0.953 and an F1-Score of 0.883. However, their MCC scores were lower in contrast to ET, RF, and DT due to their recall scores. 

Comparing the best-performing classifiers ET and RF to the baseline models shows that they perform significantly higher, except for DT, which only performed significantly worse in terms of log-loss. The t.ex-Graph classifiers display lower performance across all metrics. 

\subsection{Cross-browser and Longitudinal Performance}
Figure~\ref{fig:spiderplot} presents the cross-browser performance and longitudinal performance of our  $\text{Chrome}_{22}$-trained classifiers. We focus on three selected classifiers: one baseline model, our best performing model, and one gradient boosting model.  

For ET and XGBoost, $\text{Chrome}_{22}$ shows high performance across most metrics, which is expected as the models would be most adapted to the data they were trained on. The performance on $\text{Chrome}_{23}$, $\text{Firefox}_{22}$, and $\text{Brave}_{22}$ varies, with a general trend of decreasing performance, indicating potential overfitting to the training data or a lack of generalizability to other datasets. The former should not be the case, as we observe less than a 5\% performance drop between the in-distribution training and test set. Furthermore, we split the in-distribution data before the pre-processing, thus mitigating any data leakage and applied \textit{Repeated Stratified K-Fold} cross-validation with five repeats and five folds. As shown in the previous sections, the datasets are different and trackers have varying structures, which is why we perform a cross-browser evaluation to test the generalizability. 


Overall, the results depicted in Figure~\ref{fig:spiderplot} indicate that, while all models perform well on the dataset they were trained on, their ability to generalize to other data varies. ET exhibits the most consistent performance across different browser datasets, indicating better generalization but it had a lower log-loss score for $\text{Firefox}_{22}$ compared to the two. 
The figure underscores the importance of cross-browser testing to ensure that models are not only tuned to a specific dataset but also maintain their performance across diverse scenarios. 

The Brave results show a high accuracy metric for the overall performance, but the F1-Score and MCC are low and represent a more realistic performance, which is expected due to the considerable differences between the Chrome and Brave data. 

Our longitudinal performance analysis indicates a slight decline in the efficacy of the classifiers when transitioning from the $\text{Chrome}_{22}$ to the $\text{Chrome}_{23}$ dataset. 
We can observe a decrease in metrics such as AUC-ROC, Accuracy, and others.
Notably, the MCC score between ET and XGBoost are getting similar. A trend which we can observe across most classifiers for $\text{Chrome}_{23}$ but also $\text{Firefox}_{22}$. 
However, the MCC scores of the gradient boosting models did not drop significantly more between $\text{Chrome}_{22}$ and $\text{Chrome}_{23}$, e.g., XGBoost went down from 0.838 to 0.823, whereas the ET suffered from 0.903 to 0.851. 
This suggests that while all models are affected by evolving data characteristics, ensemble methods, particularly some gradient boosting models, may offer a more stable predictive performance in longitudinal settings in contrast to the ET and RF. Nevertheless, ET and RF had better or similar performance compared to XGBoost. 
Applying isotonic calibration has only minor improvements compared to the uncalibrated models, as highlighted by Figure~\ref{fig:calib}. The exceptions are GNB and AdaBoost, where estimates have improved, e.g., the uncalibrated AdaBoost initially centered most predictions around 0.5. 

\subsection{Feature Importance}
\begin{figure}[htbp] 
    \centering
    \includegraphics[width=1\columnwidth]{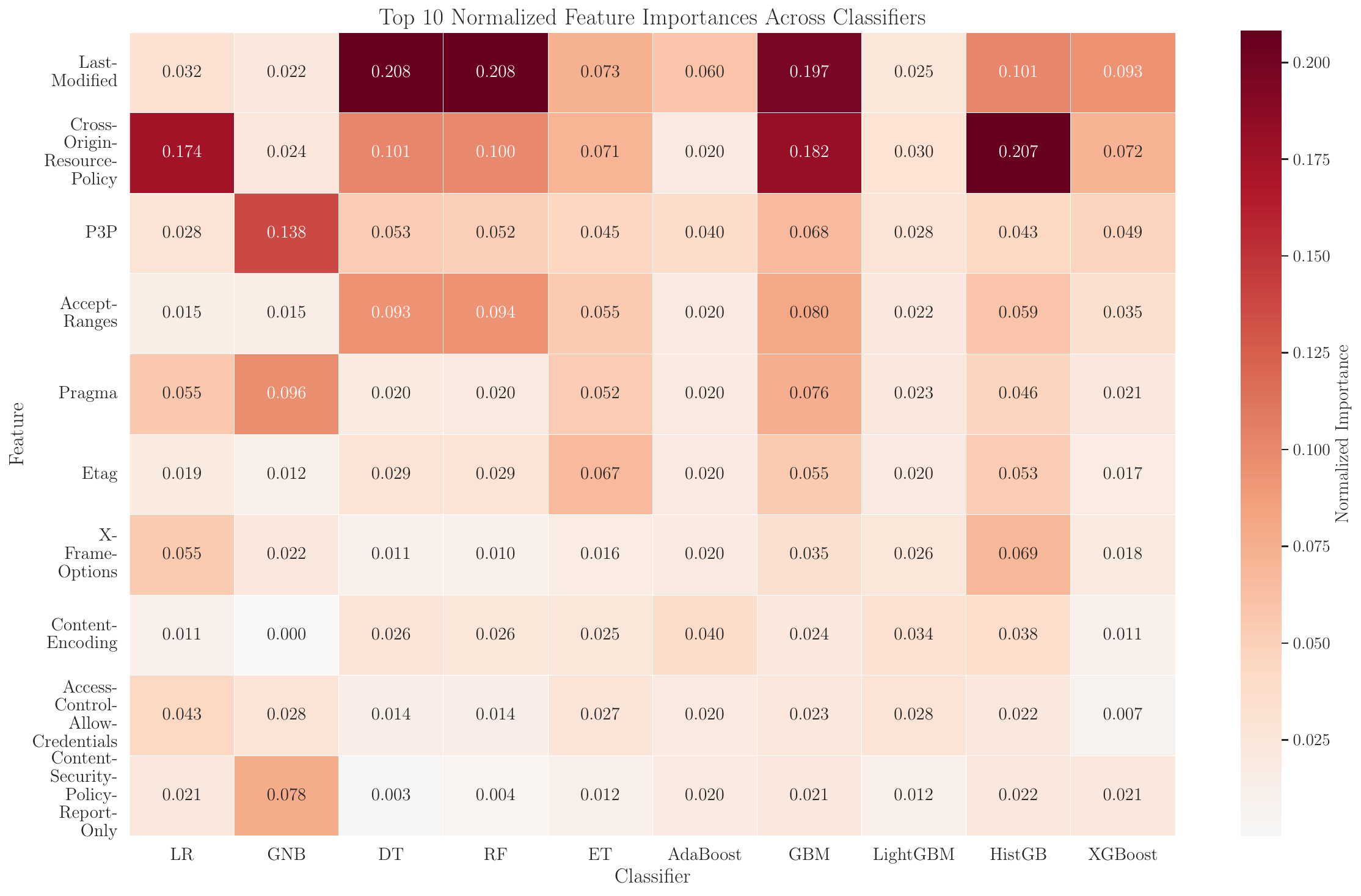}
    \caption{The top ten features are presented in order of importance across different classifiers based on the in-distribution test set. The heatmap illustrates the normalized importance values for each feature and classifier. We calculated \textit{Permutation feature importance} for LR and GNB but not for the others as we do not have any high cardinality features.} 
    \label{fig:feature_importance_top10}
    \Description[Feature importance of Chrome-based classifiers using HTTP response headers.]{The top ten features are presented in order of importance across different classifiers based on the in-distribution test set. The heatmap illustrates the normalized importance values for each feature and classifier. We calculated \textit{Permutation feature importance} for LR and GNB but not for the others as we do not have any high cardinality features.}
\end{figure}

As shown in Figure~\ref{fig:feature_importance_top10}, the features \textit{Cross-Origin-Resource-Policy} and \textit{Last-Modified} were the important features overall, followed by \textit{Accept-Ranges} and \textit{P3P}. 
These four headers have distinct purposes in HTTP communication, except for \textit{P3P} which is deprecated nowadays~\footnote{\url{https://www.w3.org/TR/P3P/}}. 
Although we plot the top ten features, the majority of them have low importance values below 0.1. DT, RF, and GBM share similar values, whereas the best-performing model, ET, has no features with a significantly higher value than the others. 

Iqbal et al. \cite{khaleesi}, who also used response header data for their classification as discussed in Section~\ref{cha2:ml}, calculated the \textit{information gain} for each of their eight response-based features. This is only partially comparable to feature importance but enables two key comparisons: (i) \textit{ETag} and especially \textit{P3P} were important features as well, and (ii) they highlighted the importance of the \textit{Content-Length}, \textit{Content-Type}, and \textit{Set-Cookie} headers, although they did not rank among the top 10 features in our classifiers.


\subsection{HTTP Request Header Performance}

\begin{figure}[htbp] 
    \centering
    \includegraphics[width=1\columnwidth]{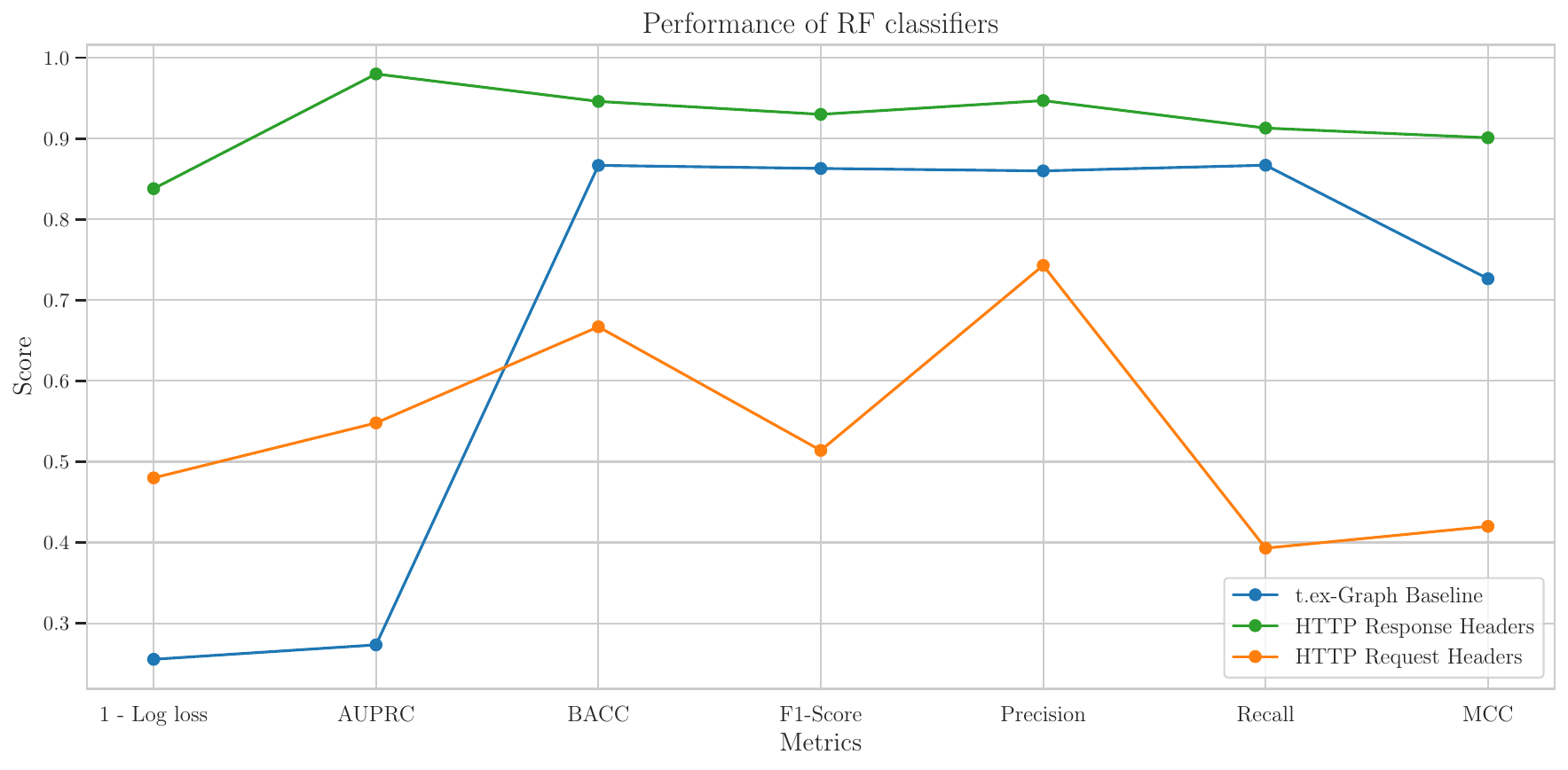}
    \caption{Comparison between our response-based, request-based, and T.EX-Graph's RF classifier. The response-based classifier outperforms its counterparts across all evaluated metrics.}
    \label{fig:request_performance}
    \Description[The RF classifier outperforms the baselines.]{Comparison between our response-based, request-based, and T.EX-Graph's RF classifier. The response-based classifier outperforms its counterparts across all evaluated metrics.}
\end{figure}

Lastly, we compare the performance of our response header-based approach to classifiers that were trained on request headers. Comparing the request-based classifiers internally, we observe performance differences similar to those of their response-based counterparts. Overall, however, the request-based classifiers perform worse, as shown in Figure~\ref{fig:request_performance}. Their AUPRC, F1-Score, and MCC values are close to 0.5 or less, which means that their prediction reliability is low. 

We also observe a pattern similar to the response-based classifiers regarding cross-browser and longitudinal performance. Over time, the classifiers perform slightly worse, e.g., with the MCC for the ET going down from 0.42 to 0.404. Regarding $\text{Firefox}_{22}$ data, the precision improved but the recall decreased. Nevertheless, the MCC is only 0.213, and the AUPRC is 0.458 for the ET, indicating that the classifier has no discriminative power. In the case of Brave, the classifiers are essentially ineffective and exhibit a low F1-score and MCC.  

Among the most important features across the majority of classifiers are \textit{X-Client-Data}, \textit{Content-Type}, \textit{Origin}, and \textit{Cookie}. The first two are especially relevant, with values ranging from 0.2 to 0.566, excluding GNB, AdaBoost, and LightGBM. Although \textit{Content-Type} is both a request and response header, it was not among the top-ten most important features for the response classifiers. The most important feature for the request-based classifiers, \textit{X-Client-Data}, is used by Chrome to send an identifier to certain Alphabet-affiliated domains, e.g., \textit{doubleclick.net} or \textit{googleadservices.com}~\footnote{\url{https://github.com/chromium/chromium/blob/master/components/variations/net/variations_http_headers.cc}}, which are prevalent in our datasets. We observe similar importance values when we use an out-of-distribution dataset like $\text{Firefox}_{22}$.

\section{Discussion}\label{sec:discussion}


Tracking affects users across all three browsers but varies to a certain degree, e.g.,\ Brave had the least amount of tracking-related HTTP messages compared to Chrome and Firefox. This can be attributed to Brave's extensive built-in privacy protection mechanisms. 



\subsection{Experimental Results}
First and foremost, the application of binarization resulted in surprisingly good results, i.e.,\ the mere presence of a header value is sufficient and extensive feature engineering is not required from a tracker detection standpoint. 
Our approach works well in this imbalanced setting, although it might be problematic for some classification algorithms without additional measures, such as setting class weights or rebalancing the class distribution. Additional measures include using ML models, which handle imbalanced datasets well \cite{JUEZGIL2021107447}, or anomaly detection \cite{Park2018AnomalyDF}. Thus, although this has not been considered in our pipeline, it might be interesting to explore how well these measures may improve the performance of our proposed approach. 

Our cross-browser approach reveals three things: firstly, the $\text{Chrome}_{22}$ dataset already captures most of the tracker structure. Secondly, due to the resulting effectiveness of a $\text{Chrome}_{22}$-trained classifier on out-of-distribution data, such as $\text{Firefox}_{22}$, it may not be necessary to re-train classifiers for these datasets. 
Thirdly, applying our approach to browsers like Brave that filter web tracking by default has a strong negative performance impact because the classifiers trained on unfiltered data are not able to effectively classify the modified structure of trackers in the filtered data. 

The results of our clustering shown in Figure~\ref{fig:tsne-clustering} reveal that trackers are more dispersed than non-trackers, which might be an argument for browser-specific models, as these might be able to better distinguish between trackers. Another challenge is the overlap between trackers and non-trackers, which further impedes correct classification. Feature engineering taking the header values into account might be a valuable approach to reduce the overlap, as header values can differ between trackers and non-trackers (see Section~\ref{sec:data-analysis}). A data-driven and large-scale in-depth analysis of header values and tracker structures across browsers is needed to explore this approach further.


Our comparison of request- and response-based classifiers revealed that request-based classifiers perform worse overall, which might be attributed to their comparatively small feature space, i.e., a smaller number of unique headers. This weaker performance suggests that request headers are less informative for the detection of web trackers than response headers. Nevertheless, the request-based classifiers correctly identify some trackers that the response-based classifiers missed, and vice-versa, as highlighted by the example of ET in Table~\ref{table:results-overview}. Thus, future work could explore combinations of both approaches as part of a multi-modal ensemble model.

\begin{table}[htbp]
    \centering
    \small
    \begin{tabular}{@{}ccccc@{}}
        \toprule
        \multicolumn{2}{c}{\textbf{Predicted Label by ET}} & \multicolumn{1}{c}{\textbf{True Label}} & \textbf{Count} & \textbf{Total} \\
        \cmidrule(r){1-2} 
        \thead{Response} & \thead{Request} & & & \\
        \midrule
        NT & NT & T & 3777 & 103357 \\
        NT & NT & NT & 99580 & 103357 \\
        T & T & NT & 404 & 7826 \\
        T & T & T & 7422 & 7826 \\
        NT & T & T & 729 & 19257 \\
        NT & T & NT & 18528 & 19257 \\
        T & NT & NT & 1991 & 41304 \\
        T & NT & T & 39313 & 41304 \\
        \bottomrule
    \end{tabular}
    \\[6pt]
    \caption{Comparison of the predicted and true labels for the request- and response-based ET classifier based on the in-distribution test set.} 
    \label{table:results-overview}
\end{table}

\subsection{Potential Evasion Approaches}
Spoofing of header values to evade our approach should not be a problem, as the values are not taken into account. 
However, this limits the models' explainability in the sense that we lose the actual header values that might provide us with relevant information to understand trackers and whether some headers are required by trackers or not. 

Servers can modify a response by adding or removing headers before responding to a client. The latter could impact the effectiveness of our classifiers, as they require the presence of headers.
Because analyzing the characteristics and use cases of each header in our datasets is unfeasible, we focus on the top ten headers from our feature importance analysis to assess the impact of their removal on our classifiers' ability to correctly identify trackers. 

For this, we must differentiate between deprecated and actively used headers. The former may be removed by servers at any time and might prevent a positive detection by our classifiers. Examples include the \textit{Pragma} header, which provides caching directives and has been superseded by the \textit{Cache-Control} header, and the \textit{P3P} header representing privacy practices. 

Headers belonging to the latter category cannot be removed as easily. Security headers, such as \textit{X-Frame-Options}, \textit{Cross-Origin-Resource-Policy}, and \textit{Content-Security-Policy-Report-Only}, are used to defend against various attacks like click-jacking, cross-site script inclusion attacks, or cross-site request forgery. Of equal importance are headers related to Cross-Origin-Resource-Sharing, such as \textit{Access-Control-Allow-Credentials} that allows the inclusion of, e.g., credentials for HTTP authentication or cookies in cross-origin requests. Other vital headers include \textit{Last-Modified}, \textit{Accept-Ranges}, \textit{Etag}, and \textit{Content-Encoding}, which are relevant for compression, caching, or downloads. Excluding any of the aforementioned headers would have consequences leading to decreased security, performance, or functionality. Therefore, it would not be in the interest of web services to exclude these headers.

Siby et al.~\cite{siby} argued that content features, such as URLs, are more vulnerable to obfuscation than structural features. This assertion extends to the content of headers as well. However, our classifiers are not concerned with the content of headers, but rather with their mere presence. Malicious actors could employ two different techniques in this case: (i) the addition of headers and (ii) the removal of headers -- both could be done intentionally or randomly. The complete removal, however, could potentially disrupt the functionality of trackers or non-trackers. 
The presented classifiers could be further enhanced through the integration or the extension of structural and \textit{flow} features. In the case of Siby et al., the inclusion of these features contributed to the robustness of \textsc{A\textsc{d}Graph}. 

The authors~\cite{siby} developed a threat model and considered various techniques that trackers might utilize to evade detection. Although it is conceivable that an individual tracker could obfuscate its identity by employing a particular set of response headers to mimic non-trackers, if all trackers were to adopt a uniform set of headers or employ a similar strategy, they would collectively form a distinct cluster as illustrated in Figure~\ref{fig:tsne-clustering}, thereby differentiating themselves from non-trackers. 

Furthermore, high coordination between trackers would be necessary to employ said strategies but due to the heterogeneity of trackers and their owners, this level of coordination is unlikely. Nevertheless, a large impact could be achieved if the owner's trackers are highly prevalent such as Alphabet's. 

Lastly, the similarity between trackers and non-trackers with respect to their feature vectors is already notably, emphasizing the classifier's effectiveness in differentiating between these two groups. This suggests that the classifier is already highly proficient in identifying and separating trackers from non-trackers.

\subsection{Deployment Scenarios}
Utilizing the models for automated web tracker detection in real-life applications is possible in two scenarios. First, HTTP response information is unavailable in an online (or live) detection scenario to block communication with web trackers. In this scenario, our model can enable or support reinforced learning of a complementary model on whose prediction blocking of HTTP communication is decided. This process, e.g., may imply manual inspection of HTTP communication where the predictions of our model are considered in cases where the primary model is wrong or indecisive. Second, as filter lists are still fundamental for web privacy protection, the manual process of generating filter rules can be optimized or outright replaced by utilizing our model. In this scenario, existing datasets or data from volunteers containing HTTP communication are used to identify web tracking requests. Filter rules for those can be dynamically generated and appended to conventional filter lists, which are already used.

\subsection{Limitations of Using Filter Lists}
As one of the oldest methods for blocking trackers~\cite{mazel}, filter lists have proven to be reliable and effective. However, they have to be updated constantly, which is commonly done manually. A longitudinal study of nine years by Alrizah et al.~\cite{alrizah} showed that human errors and delayed updates are a root cause for lower performance. Chen et al.~\cite{chen} highlighted methods to evade blockers and filter lists: (i) changing URLs, (ii) embedding code directly into the website with the \textit{script}-tag, and (iii) combining functional and tracking code in one file. In summary, using filter lists for our ground truth is not without problems and may impact the evaluation of our classifiers. Nevertheless, they present a popular ground truth in research to label datasets for training ML models.

\section{Conclusion}\label{sec:conclusion}
In this work, we train $\text{Chrome}_{22}$-based classifiers on a subset of binarized response headers to evaluate their effectiveness and usefulness in web tracker detection while retaining the natural imbalance of trackers. 

We observe that response headers represent characteristics of web trackers which were previously under-explored by past research. They can be used to differentiate well between trackers and non-trackers and provide additional information on the nature of tracking-related entities. Our data-driven approach combined with binarization reveals new and confirms previously identified features. Applying the same methodology to request headers results in a weaker performance, thus strengthening the feasibility of using response headers for web tracker detection. 

Our two best classifiers RF and ET achieve scores ranging from 0.9 to 0.98. 
In addition, we measure the cross-browser effectiveness, showing that multiple $\text{Chrome}_{22}$-based classifiers achieve high AUPRC but moderate MCC values for $\text{Firefox}_{22}$. 
However, we also observe large performance losses for $\text{Brave}_{22}$. 
The longitudinal performance of the selected classifiers deteriorates only by small margins, which highlights their potential due to the changing capabilities of trackers \cite{fouadPixel, bujlow, 1-million-analysis}. Still, periodic re-training might be necessary if the performance decreases significantly. 

In future research, we plan to expand on our approach of leveraging response headers for web tracker detection by considering header values in feature engineering and exploring the viability of deep-learning approaches based on response header data. Additionally, we intend to integrate our approach with existing solutions to develop a holistic web tracking detection system. Moreover, headers represent only a part of an HTTP message, and research has shown that the message bodies are also relevant when it comes to understanding and identifying trackers \cite{li, guarino}. Emerging technologies like Large Language Models could be used to analyze bodies in an efficient and meaningful way. 



\begin{acks}
We thank our anonymous reviewers for their valuable feedback. 
Funded by the European Union -- NextGenerationEU and the German Federal Ministry of Economics and Climate Protection (BMWK) in the research project "Werk 4.0" and is supervised by the project sponsor VDI Technologiezentrum GmbH [grant number 13IK022K]. The authors used ChatGPT4 to revise the title of this paper by asking it to make the title more interesting.  
\end{acks}

\bibliographystyle{ACM-Reference-Format}
\bibliography{main}


\begin{thebibliography}{51}


\ifx \showCODEN    \undefined \def \showCODEN     #1{\unskip}     \fi
\ifx \showDOI      \undefined \def \showDOI       #1{#1}\fi
\ifx \showISBNx    \undefined \def \showISBNx     #1{\unskip}     \fi
\ifx \showISBNxiii \undefined \def \showISBNxiii  #1{\unskip}     \fi
\ifx \showISSN     \undefined \def \showISSN      #1{\unskip}     \fi
\ifx \showLCCN     \undefined \def \showLCCN      #1{\unskip}     \fi
\ifx \shownote     \undefined \def \shownote      #1{#1}          \fi
\ifx \showarticletitle \undefined \def \showarticletitle #1{#1}   \fi
\ifx \showURL      \undefined \def \showURL       {\relax}        \fi
\providecommand\bibfield[2]{#2}
\providecommand\bibinfo[2]{#2}
\providecommand\natexlab[1]{#1}
\providecommand\showeprint[2][]{arXiv:#2}

\bibitem[Alrizah et~al\mbox{.}(2019)]%
        {alrizah}
\bibfield{author}{\bibinfo{person}{Mshabab Alrizah}, \bibinfo{person}{Sencun Zhu}, \bibinfo{person}{Xinyu Xing}, {and} \bibinfo{person}{Gang Wang}.} \bibinfo{year}{2019}\natexlab{}.
\newblock \showarticletitle{Errors, Misunderstandings, and Attacks: Analyzing the Crowdsourcing Process of Ad-blocking Systems}, In \bibinfo{booktitle}{Proceedings of the Internet Measurement Conference} (Amsterdam, Netherlands).
\newblock \bibinfo{journal}{\emph{Proceedings of the Internet Measurement Conference}}, \bibinfo{pages}{230–244}.
\newblock


\bibitem[Bahrami et~al\mbox{.}(2021)]%
        {bahrami}
\bibfield{author}{\bibinfo{person}{Pouneh~Nikkhah Bahrami}, \bibinfo{person}{Umar Iqbal}, {and} \bibinfo{person}{Zubair Shafiq}.} \bibinfo{year}{2021}\natexlab{}.
\newblock \showarticletitle{FP-Radar: Longitudinal Measurement and Early Detection of Browser Fingerprinting}.
\newblock \bibinfo{journal}{\emph{Proceedings on Privacy Enhancing Technologies}}  \bibinfo{volume}{2022} (\bibinfo{year}{2021}), \bibinfo{pages}{557 -- 577}.
\newblock


\bibitem[Bau et~al\mbox{.}(2013)]%
        {bau}
\bibfield{author}{\bibinfo{person}{Jason Bau}, \bibinfo{person}{Jonathan~R. Mayer}, \bibinfo{person}{Hristo~S. Paskov}, {and} \bibinfo{person}{John~C. Mitchell}.} \bibinfo{year}{2013}\natexlab{}.
\newblock \showarticletitle{A Promising Direction for Web Tracking Countermeasures}.
\newblock \bibinfo{journal}{\emph{Proceedings of W2SP}} (\bibinfo{year}{2013}).
\newblock


\bibitem[Bhagavatula et~al\mbox{.}(2014)]%
        {bhagavatula}
\bibfield{author}{\bibinfo{person}{Sruti Bhagavatula}, \bibinfo{person}{Christopher Dunn}, \bibinfo{person}{Chris Kanich}, \bibinfo{person}{Minaxi Gupta}, {and} \bibinfo{person}{Brian Ziebart}.} \bibinfo{year}{2014}\natexlab{}.
\newblock \showarticletitle{Leveraging Machine Learning to Improve Unwanted Resource Filtering}. In \bibinfo{booktitle}{\emph{Proceedings of the 2014 Workshop on Artificial Intelligent and Security Workshop}} (Scottsdale, Arizona, USA) \emph{(\bibinfo{series}{AISec '14})}. \bibinfo{publisher}{Association for Computing Machinery}, \bibinfo{address}{New York, NY, USA}, \bibinfo{pages}{95–102}.
\newblock
\showISBNx{9781450331531}


\bibitem[Boughorbel et~al\mbox{.}(2017)]%
        {mcc2}
\bibfield{author}{\bibinfo{person}{Sabri Boughorbel}, \bibinfo{person}{Fethi Jarray}, {and} \bibinfo{person}{Mohammed El-Anbari}.} \bibinfo{year}{2017}\natexlab{}.
\newblock \showarticletitle{Optimal classifier for imbalanced data using Matthews Correlation Coefficient metric}.
\newblock \bibinfo{journal}{\emph{PLoS ONE}}  \bibinfo{volume}{12} (\bibinfo{year}{2017}).
\newblock


\bibitem[Brodersen et~al\mbox{.}(2010)]%
        {broderson}
\bibfield{author}{\bibinfo{person}{Kay~Henning Brodersen}, \bibinfo{person}{Cheng~Soon Ong}, \bibinfo{person}{Klaas~Enno Stephan}, {and} \bibinfo{person}{Joachim~M. Buhmann}.} \bibinfo{year}{2010}\natexlab{}.
\newblock \showarticletitle{The Balanced Accuracy and Its Posterior Distribution}. In \bibinfo{booktitle}{\emph{2010 20th International Conference on Pattern Recognition}}. \bibinfo{pages}{3121--3124}.
\newblock


\bibitem[Bujlow et~al\mbox{.}(2017)]%
        {bujlow}
\bibfield{author}{\bibinfo{person}{Tomasz Bujlow}, \bibinfo{person}{Valentín Carela-Español}, \bibinfo{person}{Josep Solé-Pareta}, {and} \bibinfo{person}{Pere Barlet-Ros}.} \bibinfo{year}{2017}\natexlab{}.
\newblock \showarticletitle{A Survey on Web Tracking: Mechanisms, Implications, and Defenses}.
\newblock \bibinfo{journal}{\emph{Proc. IEEE}} \bibinfo{volume}{105}, \bibinfo{number}{8} (\bibinfo{year}{2017}), \bibinfo{pages}{1476--1510}.
\newblock


\bibitem[Castell-Uroz et~al\mbox{.}(2021)]%
        {castell}
\bibfield{author}{\bibinfo{person}{Ismael Castell-Uroz}, \bibinfo{person}{Josep Solé-Pareta}, {and} \bibinfo{person}{Pere Barlet-Ros}.} \bibinfo{year}{2021}\natexlab{}.
\newblock \showarticletitle{TrackSign: Guided Web Tracking Discovery}. In \bibinfo{booktitle}{\emph{IEEE INFOCOM 2021 - IEEE Conference on Computer Communications}}. \bibinfo{pages}{1--10}.
\newblock


\bibitem[Chanchary and Chiasson(2015)]%
        {chanchary}
\bibfield{author}{\bibinfo{person}{Farah~Habib Chanchary} {and} \bibinfo{person}{Sonia Chiasson}.} \bibinfo{year}{2015}\natexlab{}.
\newblock \showarticletitle{User Perceptions of Sharing, Advertising, and Tracking}. In \bibinfo{booktitle}{\emph{Symposium On Usable Privacy and Security}}.
\newblock


\bibitem[Chen et~al\mbox{.}(2021)]%
        {chen}
\bibfield{author}{\bibinfo{person}{Quan Chen}, \bibinfo{person}{Peter Snyder}, \bibinfo{person}{Ben Livshits}, {and} \bibinfo{person}{Alexandros Kapravelos}.} \bibinfo{year}{2021}\natexlab{}.
\newblock \showarticletitle{Detecting Filter List Evasion with Event-Loop-Turn Granularity JavaScript Signatures}.
\newblock \bibinfo{journal}{\emph{2021 IEEE Symposium on Security and Privacy (SP)}}  \bibinfo{volume}{00} (\bibinfo{year}{2021}), \bibinfo{pages}{1715–1729}.
\newblock


\bibitem[Chicco and Jurman(2020)]%
        {chicco}
\bibfield{author}{\bibinfo{person}{Davide Chicco} {and} \bibinfo{person}{Giuseppe Jurman}.} \bibinfo{year}{2020}\natexlab{}.
\newblock \showarticletitle{The advantages of the Matthews correlation coefficient (MCC) over F1 score and accuracy in binary classification evaluation}.
\newblock \bibinfo{journal}{\emph{BMC Genomics}} \bibinfo{volume}{21}, \bibinfo{number}{1} (\bibinfo{year}{2020}), \bibinfo{pages}{6}.
\newblock


\bibitem[Cozza et~al\mbox{.}(2020)]%
        {cozza}
\bibfield{author}{\bibinfo{person}{Federico Cozza}, \bibinfo{person}{Alfonso Guarino}, \bibinfo{person}{Francesco Isernia}, \bibinfo{person}{Delfina Malandrino}, \bibinfo{person}{Antonio Rapuano}, \bibinfo{person}{Raffaele Schiavone}, {and} \bibinfo{person}{Rocco Zaccagnino}.} \bibinfo{year}{2020}\natexlab{}.
\newblock \showarticletitle{Hybrid and lightweight detection of third party tracking: Design, implementation, and evaluation}.
\newblock \bibinfo{journal}{\emph{Computer Networks}}  \bibinfo{volume}{167} (\bibinfo{year}{2020}), \bibinfo{pages}{106993}.
\newblock
\showISSN{1389-1286}


\bibitem[Demir et~al\mbox{.}(2022)]%
        {demir}
\bibfield{author}{\bibinfo{person}{Nurullah Demir}, \bibinfo{person}{Matteo Gro\ss{}e-Kampmann}, \bibinfo{person}{Tobias Urban}, \bibinfo{person}{Christian Wressnegger}, \bibinfo{person}{Thorsten Holz}, {and} \bibinfo{person}{Norbert Pohlmann}.} \bibinfo{year}{2022}\natexlab{}.
\newblock \showarticletitle{Reproducibility and Replicability of Web Measurement Studies}. In \bibinfo{booktitle}{\emph{Proceedings of the ACM Web Conference 2022}} (Virtual Event, Lyon, France) \emph{(\bibinfo{series}{WWW '22})}. \bibinfo{publisher}{Association for Computing Machinery}, \bibinfo{address}{New York, NY, USA}, \bibinfo{pages}{533–544}.
\newblock
\showISBNx{9781450390965}


\bibitem[Dietterich(2000)]%
        {ensemble}
\bibfield{author}{\bibinfo{person}{Thomas~G. Dietterich}.} \bibinfo{year}{2000}\natexlab{}.
\newblock \showarticletitle{Ensemble Methods in Machine Learning}. In \bibinfo{booktitle}{\emph{Multiple Classifier Systems}}. \bibinfo{publisher}{Springer Berlin Heidelberg}, \bibinfo{address}{Berlin, Heidelberg}, \bibinfo{pages}{1--15}.
\newblock
\showISBNx{978-3-540-45014-6}


\bibitem[Din et~al\mbox{.}(2020)]%
        {din}
\bibfield{author}{\bibinfo{person}{Zainul~Abi Din}, \bibinfo{person}{Panagiotis Tigas}, \bibinfo{person}{Samuel~T. King}, {and} \bibinfo{person}{Benjamin Livshits}.} \bibinfo{year}{2020}\natexlab{}.
\newblock \showarticletitle{PERCIVAL: Making in-Browser Perceptual Ad Blocking Practical with Deep Learning}. In \bibinfo{booktitle}{\emph{Proceedings of the 2020 USENIX Conference on Usenix Annual Technical Conference}} \emph{(\bibinfo{series}{USENIX ATC'20})}. \bibinfo{publisher}{USENIX Association}, \bibinfo{address}{USA}, Article \bibinfo{articleno}{26}, \bibinfo{numpages}{14}~pages.
\newblock
\showISBNx{978-1-939133-14-4}


\bibitem[Dudykevych and Nechypor(2016)]%
        {dudykevych}
\bibfield{author}{\bibinfo{person}{Valery Dudykevych} {and} \bibinfo{person}{Vitalii Nechypor}.} \bibinfo{year}{2016}\natexlab{}.
\newblock \showarticletitle{Detecting Third-Party User Trackers with Cookie Files}.
\newblock \bibinfo{journal}{\emph{2016 Third International Scientific-Practical Conference Problems of Infocommunications Science and Technology (PIC S\&T)}} (\bibinfo{year}{2016}), \bibinfo{pages}{78–80}.
\newblock


\bibitem[Engelmann and Lessmann(2021)]%
        {gan}
\bibfield{author}{\bibinfo{person}{Justin Engelmann} {and} \bibinfo{person}{Stefan Lessmann}.} \bibinfo{year}{2021}\natexlab{}.
\newblock \showarticletitle{Conditional Wasserstein GAN-based oversampling of tabular data for imbalanced learning}.
\newblock \bibinfo{journal}{\emph{Expert Systems with Applications}}  \bibinfo{volume}{174} (\bibinfo{year}{2021}), \bibinfo{pages}{114582}.
\newblock


\bibitem[Englehardt and Narayanan(2016)]%
        {1-million-analysis}
\bibfield{author}{\bibinfo{person}{Steven Englehardt} {and} \bibinfo{person}{Arvind Narayanan}.} \bibinfo{year}{2016}\natexlab{}.
\newblock \showarticletitle{Online Tracking: A 1-Million-Site Measurement and Analysis}. In \bibinfo{booktitle}{\emph{Proceedings of the 2016 ACM SIGSAC Conference on Computer and Communications Security}} (Vienna, Austria) \emph{(\bibinfo{series}{CCS '16})}. \bibinfo{publisher}{Association for Computing Machinery}, \bibinfo{address}{New York, NY, USA}, \bibinfo{pages}{1388–1401}.
\newblock
\showISBNx{9781450341394}


\bibitem[Fajardo et~al\mbox{.}(2021)]%
        {autoencoders}
\bibfield{author}{\bibinfo{person}{Val~Andrei Fajardo}, \bibinfo{person}{David Findlay}, \bibinfo{person}{Charu Jaiswal}, \bibinfo{person}{Xinshang Yin}, \bibinfo{person}{Roshanak Houmanfar}, \bibinfo{person}{Honglei Xie}, \bibinfo{person}{Jiaxi Liang}, \bibinfo{person}{Xichen She}, {and} \bibinfo{person}{David~B. Emerson}.} \bibinfo{year}{2021}\natexlab{}.
\newblock \showarticletitle{On oversampling imbalanced data with deep conditional generative models}.
\newblock \bibinfo{journal}{\emph{Expert Syst. Appl.}}  \bibinfo{volume}{169} (\bibinfo{year}{2021}), \bibinfo{pages}{114463}.
\newblock


\bibitem[Fouad et~al\mbox{.}(2018)]%
        {fouadPixel}
\bibfield{author}{\bibinfo{person}{Imane Fouad}, \bibinfo{person}{Nataliia Bielova}, \bibinfo{person}{Arnaud Legout}, {and} \bibinfo{person}{Natasa Sarafijanovic-Djukic}.} \bibinfo{year}{2018}\natexlab{}.
\newblock \showarticletitle{Tracking the pixels: Detecting web trackers via analyzing invisible pixels}.
\newblock \bibinfo{journal}{\emph{arXiv preprint arXiv:1812.01514}} (\bibinfo{year}{2018}).
\newblock


\bibitem[Guarino. et~al\mbox{.}(2020)]%
        {guarino}
\bibfield{author}{\bibinfo{person}{Alfonso Guarino.}, \bibinfo{person}{Delfina Malandrino.}, \bibinfo{person}{Rocco Zaccagnino.}, \bibinfo{person}{Federico Cozza.}, {and} \bibinfo{person}{Antonio Rapuano.}} \bibinfo{year}{2020}\natexlab{}.
\newblock \showarticletitle{On Analyzing Third-party Tracking via Machine Learning}. In \bibinfo{booktitle}{\emph{Proceedings of the 6th International Conference on Information Systems Security and Privacy - ICISSP,}}. INSTICC, \bibinfo{publisher}{SciTePress}, \bibinfo{pages}{532--539}.
\newblock
\showISBNx{978-989-758-399-5}
\showISSN{2184-4356}


\bibitem[Gugelmann et~al\mbox{.}(2015)]%
        {Gugelmann2015}
\bibfield{author}{\bibinfo{person}{David Gugelmann}, \bibinfo{person}{Markus Happe}, \bibinfo{person}{Bernhard Ager}, {and} \bibinfo{person}{Vincent Lenders}.} \bibinfo{year}{2015}\natexlab{}.
\newblock \showarticletitle{{An Automated Approach for Complementing Ad Blockers' Blacklists}}.
\newblock \bibinfo{journal}{\emph{Proceedings on Privacy Enhancing Technologies}} \bibinfo{volume}{2015}, \bibinfo{number}{2} (\bibinfo{year}{2015}), \bibinfo{pages}{282--298}.
\newblock


\bibitem[Hashmi et~al\mbox{.}(2019)]%
        {hashmi}
\bibfield{author}{\bibinfo{person}{Saad~Sajid Hashmi}, \bibinfo{person}{Muhammad Ikram}, {and} \bibinfo{person}{Mohamed~Ali Kaafar}.} \bibinfo{year}{2019}\natexlab{}.
\newblock \showarticletitle{A Longitudinal Analysis of Online Ad-Blocking Blacklists}.
\newblock \bibinfo{journal}{\emph{2019 IEEE 44th LCN Symposium on Emerging Topics in Networking (LCN Symposium)}}  \bibinfo{volume}{00} (\bibinfo{year}{2019}), \bibinfo{pages}{158–165}.
\newblock


\bibitem[Hernandez et~al\mbox{.}(2013)]%
        {sampling}
\bibfield{author}{\bibinfo{person}{Julio Hernandez}, \bibinfo{person}{Jes{\'u}s~Ariel Carrasco-Ochoa}, {and} \bibinfo{person}{Jos{\'e}~Francisco Mart{\'\i}nez-Trinidad}.} \bibinfo{year}{2013}\natexlab{}.
\newblock \showarticletitle{An Empirical Study of Oversampling and Undersampling for Instance Selection Methods on Imbalance Datasets}. In \bibinfo{booktitle}{\emph{Progress in Pattern Recognition, Image Analysis, Computer Vision, and Applications}}, \bibfield{editor}{\bibinfo{person}{Jos{\'e} Ruiz-Shulcloper} {and} \bibinfo{person}{Gabriella Sanniti~di Baja}} (Eds.). \bibinfo{publisher}{Springer Berlin Heidelberg}, \bibinfo{address}{Berlin, Heidelberg}, \bibinfo{pages}{262--269}.
\newblock


\bibitem[Iqbal et~al\mbox{.}(2021)]%
        {iqbalML}
\bibfield{author}{\bibinfo{person}{Umar Iqbal}, \bibinfo{person}{Steven Englehardt}, {and} \bibinfo{person}{Zubair Shafiq}.} \bibinfo{year}{2021}\natexlab{}.
\newblock \showarticletitle{Fingerprinting the Fingerprinters: Learning to Detect Browser Fingerprinting Behaviors}. In \bibinfo{booktitle}{\emph{2021 IEEE Symposium on Security and Privacy (SP)}}. \bibinfo{pages}{1143--1161}.
\newblock
\urldef\tempurl%
\url{https://doi.org/10.1109/SP40001.2021.00017}
\showDOI{\tempurl}


\bibitem[Iqbal et~al\mbox{.}(2022)]%
        {khaleesi}
\bibfield{author}{\bibinfo{person}{Umar Iqbal}, \bibinfo{person}{Charlie Wolfe}, \bibinfo{person}{Charles Nguyen}, \bibinfo{person}{Steven Englehardt}, {and} \bibinfo{person}{Zubair Shafiq}.} \bibinfo{year}{2022}\natexlab{}.
\newblock \showarticletitle{Khaleesi: Breaker of Advertising and Tracking Request Chains}. In \bibinfo{booktitle}{\emph{31st USENIX Security Symposium (USENIX Security 22)}}. \bibinfo{publisher}{USENIX Association}, \bibinfo{address}{Boston, MA}, \bibinfo{pages}{2911--2928}.
\newblock
\showISBNx{978-1-939133-31-1}


\bibitem[Juez-Gil et~al\mbox{.}(2021)]%
        {JUEZGIL2021107447}
\bibfield{author}{\bibinfo{person}{Mario Juez-Gil}, \bibinfo{person}{{\'A}lvar Arnaiz-Gonz{\'a}lez}, \bibinfo{person}{Juan~J. Rodr{\'\i}guez}, {and} \bibinfo{person}{C{\'e}sar Garc{\'\i}a-Osorio}.} \bibinfo{year}{2021}\natexlab{}.
\newblock \showarticletitle{Experimental evaluation of ensemble classifiers for imbalance in Big Data}.
\newblock \bibinfo{journal}{\emph{Applied Soft Computing}}  \bibinfo{volume}{108} (\bibinfo{year}{2021}), \bibinfo{pages}{107447}.
\newblock


\bibitem[Kalavri et~al\mbox{.}(2016)]%
        {kalavri}
\bibfield{author}{\bibinfo{person}{Vasiliki Kalavri}, \bibinfo{person}{Jeremy Blackburn}, \bibinfo{person}{Matteo Varvello}, {and} \bibinfo{person}{Konstantina Papagiannaki}.} \bibinfo{year}{2016}\natexlab{}.
\newblock \showarticletitle{Like a Pack of Wolves: Community Structure of Web Trackers}. In \bibinfo{booktitle}{\emph{Passive and Active Measurement}}, \bibfield{editor}{\bibinfo{person}{Thomas Karagiannis} {and} \bibinfo{person}{Xenofontas Dimitropoulos}} (Eds.). \bibinfo{publisher}{Springer International Publishing}, \bibinfo{address}{Cham}, \bibinfo{pages}{42--54}.
\newblock


\bibitem[Laughter et~al\mbox{.}(2021)]%
        {laughter}
\bibfield{author}{\bibinfo{person}{Ashley Laughter}, \bibinfo{person}{Safwan Omari}, \bibinfo{person}{Piotr Szczurek}, {and} \bibinfo{person}{Jason Perry}.} \bibinfo{year}{2021}\natexlab{}.
\newblock \showarticletitle{Detection of Malicious HTTP Requests Using Header and URL Features}. In \bibinfo{booktitle}{\emph{Proceedings of the Future Technologies Conference (FTC) 2020, Volume 2}}, \bibfield{editor}{\bibinfo{person}{Kohei Arai}, \bibinfo{person}{Supriya Kapoor}, {and} \bibinfo{person}{Rahul Bhatia}} (Eds.). \bibinfo{publisher}{Springer International Publishing}, \bibinfo{address}{Cham}, \bibinfo{pages}{449--468}.
\newblock


\bibitem[Li et~al\mbox{.}(2015)]%
        {li}
\bibfield{author}{\bibinfo{person}{Tai-Ching Li}, \bibinfo{person}{Huy Hang}, \bibinfo{person}{Michalis Faloutsos}, {and} \bibinfo{person}{Petros Efstathopoulos}.} \bibinfo{year}{2015}\natexlab{}.
\newblock \showarticletitle{TrackAdvisor: Taking Back Browsing Privacy from Third-Party Trackers}. In \bibinfo{booktitle}{\emph{Passive and Active Measurement}}, \bibfield{editor}{\bibinfo{person}{Jelena Mirkovic} {and} \bibinfo{person}{Yong Liu}} (Eds.). \bibinfo{publisher}{Springer International Publishing}, \bibinfo{address}{Cham}, \bibinfo{pages}{277--289}.
\newblock
\showISBNx{978-3-319-15509-8}


\bibitem[Matthews(1975)]%
        {mcc}
\bibfield{author}{\bibinfo{person}{B.W. Matthews}.} \bibinfo{year}{1975}\natexlab{}.
\newblock \showarticletitle{Comparison of the predicted and observed secondary structure of T4 phage lysozyme}.
\newblock \bibinfo{journal}{\emph{Biochimica et Biophysica Acta (BBA) - Protein Structure}} \bibinfo{volume}{405}, \bibinfo{number}{2} (\bibinfo{year}{1975}), \bibinfo{pages}{442--451}.
\newblock


\bibitem[Mazel et~al\mbox{.}(2019)]%
        {mazel}
\bibfield{author}{\bibinfo{person}{Johan Mazel}, \bibinfo{person}{Richard Garnier}, {and} \bibinfo{person}{Kensuke Fukuda}.} \bibinfo{year}{2019}\natexlab{}.
\newblock \showarticletitle{A comparison of web privacy protection techniques}.
\newblock \bibinfo{journal}{\emph{Computer Communications}}  \bibinfo{volume}{144} (\bibinfo{year}{2019}), \bibinfo{pages}{162--174}.
\newblock
\showISSN{0140-3664}


\bibitem[McGahagan et~al\mbox{.}(2019)]%
        {mcGahagan}
\bibfield{author}{\bibinfo{person}{John McGahagan}, \bibinfo{person}{Darshan Bhansali}, \bibinfo{person}{Margaret Gratian}, {and} \bibinfo{person}{Michel Cukier}.} \bibinfo{year}{2019}\natexlab{}.
\newblock \showarticletitle{A Comprehensive Evaluation of HTTP Header Features for Detecting Malicious Websites}. In \bibinfo{booktitle}{\emph{2019 15th European Dependable Computing Conference (EDCC)}}. \bibinfo{pages}{75--82}.
\newblock


\bibitem[Melicher et~al\mbox{.}(2016)]%
        {melicher}
\bibfield{author}{\bibinfo{person}{William Melicher}, \bibinfo{person}{Mahmood Sharif}, \bibinfo{person}{Joshua Tan}, \bibinfo{person}{Lujo Bauer}, \bibinfo{person}{Mihai Christodorescu}, {and} \bibinfo{person}{Pedro~Giovanni Leon}.} \bibinfo{year}{2016}\natexlab{}.
\newblock \showarticletitle{(Do Not) Track Me Sometimes: Users’ Contextual Preferences for Web Tracking}.
\newblock \bibinfo{journal}{\emph{Proceedings on Privacy Enhancing Technologies}}  \bibinfo{volume}{2016} (\bibinfo{year}{2016}), \bibinfo{pages}{135 -- 154}.
\newblock


\bibitem[Metwalley et~al\mbox{.}(2015)]%
        {metwalley}
\bibfield{author}{\bibinfo{person}{Hassan Metwalley}, \bibinfo{person}{Stefano Traverso}, {and} \bibinfo{person}{Marco Mellia}.} \bibinfo{year}{2015}\natexlab{}.
\newblock \showarticletitle{Unsupervised Detection of Web Trackers}. In \bibinfo{booktitle}{\emph{2015 IEEE Global Communications Conference (GLOBECOM)}}. \bibinfo{pages}{1--6}.
\newblock


\bibitem[Niculescu-Mizil and Caruana(2005)]%
        {calibration}
\bibfield{author}{\bibinfo{person}{Alexandru Niculescu-Mizil} {and} \bibinfo{person}{Rich Caruana}.} \bibinfo{year}{2005}\natexlab{}.
\newblock \showarticletitle{Predicting Good Probabilities with Supervised Learning}. In \bibinfo{booktitle}{\emph{Proceedings of the 22nd International Conference on Machine Learning}} (Bonn, Germany) \emph{(\bibinfo{series}{ICML '05})}. \bibinfo{publisher}{Association for Computing Machinery}, \bibinfo{address}{New York, NY, USA}, \bibinfo{pages}{625–632}.
\newblock


\bibitem[Park et~al\mbox{.}(2018)]%
        {Park2018AnomalyDF}
\bibfield{author}{\bibinfo{person}{Seungyoung Park}, \bibinfo{person}{Myungjin Kim}, {and} \bibinfo{person}{Seokwoo Lee}.} \bibinfo{year}{2018}\natexlab{}.
\newblock \showarticletitle{Anomaly Detection for HTTP Using Convolutional Autoencoders}.
\newblock \bibinfo{journal}{\emph{IEEE Access}}  \bibinfo{volume}{6} (\bibinfo{year}{2018}), \bibinfo{pages}{70884--70901}.
\newblock


\bibitem[Pochat et~al\mbox{.}(2018)]%
        {pochat}
\bibfield{author}{\bibinfo{person}{Victor~Le Pochat}, \bibinfo{person}{Tom van Goethem}, \bibinfo{person}{Samaneh Tajalizadehkhoob}, \bibinfo{person}{Maciej Korczyński}, {and} \bibinfo{person}{Wouter Joosen}.} \bibinfo{year}{2018}\natexlab{}.
\newblock \showarticletitle{Tranco: A Research-Oriented Top Sites Ranking Hardened Against Manipulation}.
\newblock \bibinfo{journal}{\emph{Proceedings 2019 Network and Distributed System Security Symposium}} (\bibinfo{year}{2018}).
\newblock


\bibitem[Raschke(2023a)]%
        {zenodo}
\bibfield{author}{\bibinfo{person}{Philip Raschke}.} \bibinfo{year}{2023}\natexlab{a}.
\newblock \bibinfo{booktitle}{\emph{Tranco 16-5-22 top 10K crawled with T.EX’}}.
\newblock
\urldef\tempurl%
\url{https://doi.org/10.5281/zenodo.7123945}
\showURL{%
Retrieved Jan 21, 2023 from \tempurl}


\bibitem[Raschke(2023b)]%
        {zenodo-new}
\bibfield{author}{\bibinfo{person}{Philip Raschke}.} \bibinfo{year}{2023}\natexlab{b}.
\newblock \bibinfo{booktitle}{\emph{Tranco 29-3-23 top 10K crawled with T.EX}}.
\newblock
\urldef\tempurl%
\url{https://doi.org/10.5281/zenodo.11555919}
\showURL{%
Retrieved Jun 12, 2024 from \tempurl}


\bibitem[Raschke and Cory(2022)]%
        {raschke2022}
\bibfield{author}{\bibinfo{person}{Philip Raschke} {and} \bibinfo{person}{Thomas Cory}.} \bibinfo{year}{2022}\natexlab{}.
\newblock \showarticletitle{Presenting a Client-based Cross-browser Web Privacy Measurement Framework for Automated Web Tracker Detection Research}.
\newblock \bibinfo{journal}{\emph{2022 3rd International Conference on Electrical Engineering and Informatics (ICon EEI)}}  \bibinfo{volume}{00} (\bibinfo{year}{2022}), \bibinfo{pages}{98–103}.
\newblock


\bibitem[Raschke et~al\mbox{.}(2023)]%
        {t.ex-graph}
\bibfield{author}{\bibinfo{person}{Philip Raschke}, \bibinfo{person}{Patrick Herbke}, {and} \bibinfo{person}{Henry Schwerdtner}.} \bibinfo{year}{2023}\natexlab{}.
\newblock \showarticletitle{t.ex-Graph: Automated Web Tracker Detection Using Centrality Metrics and Data Flow Characteristics}. In \bibinfo{booktitle}{\emph{Proceedings of the 9th International Conference on Information Systems Security and Privacy - Volume 1: ICISSP,}}. INSTICC, \bibinfo{publisher}{SciTePress}, \bibinfo{pages}{199--209}.
\newblock
\showISBNx{978-989-758-624-8}


\bibitem[Raschke et~al\mbox{.}(2019)]%
        {philipRaschke}
\bibfield{author}{\bibinfo{person}{Philip Raschke}, \bibinfo{person}{Sebastian Zickau}, \bibinfo{person}{Jacob~Leon Kr{\"o}ger}, {and} \bibinfo{person}{Axel K{\"u}pper}.} \bibinfo{year}{2019}\natexlab{}.
\newblock \showarticletitle{Towards Real-Time Web Tracking Detection with T.EX - The Transparency EXtension}. In \bibinfo{booktitle}{\emph{Privacy Technologies and Policy}}, \bibfield{editor}{\bibinfo{person}{Maurizio Naldi}, \bibinfo{person}{Giuseppe~F. Italiano}, \bibinfo{person}{Kai Rannenberg}, \bibinfo{person}{Manel Medina}, {and} \bibinfo{person}{Athena Bourka}} (Eds.). \bibinfo{publisher}{Springer International Publishing}, \bibinfo{address}{Cham}, \bibinfo{pages}{3--17}.
\newblock
\showISBNx{978-3-030-21752-5}


\bibitem[Saito and Rehmsmeier(2015)]%
        {Saito2015ThePP}
\bibfield{author}{\bibinfo{person}{Takaya Saito} {and} \bibinfo{person}{Marc Rehmsmeier}.} \bibinfo{year}{2015}\natexlab{}.
\newblock \showarticletitle{The Precision-Recall Plot Is More Informative than the ROC Plot When Evaluating Binary Classifiers on Imbalanced Datasets}.
\newblock \bibinfo{journal}{\emph{PLoS ONE}}  \bibinfo{volume}{10} (\bibinfo{year}{2015}).
\newblock


\bibitem[Santos et~al\mbox{.}(2018)]%
        {8492368}
\bibfield{author}{\bibinfo{person}{Miriam~Seoane Santos}, \bibinfo{person}{Jastin~Pompeu Soares}, \bibinfo{person}{Pedro~Henrigues Abreu}, \bibinfo{person}{Helder Araujo}, {and} \bibinfo{person}{Joao Santos}.} \bibinfo{year}{2018}\natexlab{}.
\newblock \showarticletitle{Cross-Validation for Imbalanced Datasets: Avoiding Overoptimistic and Overfitting Approaches}.
\newblock \bibinfo{journal}{\emph{IEEE Computational Intelligence Magazine}} \bibinfo{volume}{13}, \bibinfo{number}{4} (\bibinfo{year}{2018}), \bibinfo{pages}{59--76}.
\newblock


\bibitem[Siby et~al\mbox{.}(2022)]%
        {siby}
\bibfield{author}{\bibinfo{person}{Sandra Siby}, \bibinfo{person}{Umar Iqbal}, \bibinfo{person}{Steven Englehardt}, \bibinfo{person}{Zubair Shafiq}, {and} \bibinfo{person}{Carmela Troncoso}.} \bibinfo{year}{2022}\natexlab{}.
\newblock \showarticletitle{{WebGraph}: Capturing Advertising and Tracking Information Flows for Robust Blocking}. In \bibinfo{booktitle}{\emph{31st USENIX Security Symposium (USENIX Security 22)}}. \bibinfo{publisher}{USENIX Association}, \bibinfo{address}{Boston, MA}, \bibinfo{pages}{2875--2892}.
\newblock
\showISBNx{978-1-939133-31-1}


\bibitem[Steinwart and Christmann(2008)]%
        {Steinwart2008SupportVM}
\bibfield{author}{\bibinfo{person}{Ingo Steinwart} {and} \bibinfo{person}{Andreas Christmann}.} \bibinfo{year}{2008}\natexlab{}.
\newblock \bibinfo{booktitle}{\emph{Support Vector Machines} (\bibinfo{edition}{1st} ed.)}.
\newblock \bibinfo{publisher}{Springer Publishing Company, Incorporated}.
\newblock
\showISBNx{0387772413}


\bibitem[Traverso et~al\mbox{.}(2017)]%
        {traverso}
\bibfield{author}{\bibinfo{person}{Stefano Traverso}, \bibinfo{person}{Martino Trevisan}, \bibinfo{person}{Leonardo Giannantoni}, \bibinfo{person}{Marco Mellia}, {and} \bibinfo{person}{Hassan Metwalley}.} \bibinfo{year}{2017}\natexlab{}.
\newblock \showarticletitle{Benchmark and comparison of tracker-blockers: Should you trust them?}. In \bibinfo{booktitle}{\emph{2017 Network Traffic Measurement and Analysis Conference (TMA)}}. \bibinfo{publisher}{IEEE Computer Society}, \bibinfo{pages}{1--9}.
\newblock


\bibitem[Wilcox(2001)]%
        {wilcox2010fundamentals}
\bibfield{author}{\bibinfo{person}{Rand Wilcox}.} \bibinfo{year}{2001}\natexlab{}.
\newblock \bibinfo{booktitle}{\emph{Fundamentals of Modern Statistical Methods: Substantially Improving Power and Accuracy}}.
\newblock \bibinfo{publisher}{Springer}, \bibinfo{address}{New York, NY}.
\newblock
\showISBNx{978-1-4419-5524-1}


\bibitem[Wu et~al\mbox{.}(2016)]%
        {dmtTrackerDetector}
\bibfield{author}{\bibinfo{person}{Qianru Wu}, \bibinfo{person}{Qixu Liu}, \bibinfo{person}{Yuqing Zhang}, \bibinfo{person}{Peng Liu}, {and} \bibinfo{person}{Guanxing Wen}.} \bibinfo{year}{2016}\natexlab{}.
\newblock \showarticletitle{A Machine Learning Approach for Detecting Third-Party Trackers on the Web}. In \bibinfo{booktitle}{\emph{Computer Security -- ESORICS 2016}}, \bibfield{editor}{\bibinfo{person}{Ioannis Askoxylakis}, \bibinfo{person}{Sotiris Ioannidis}, \bibinfo{person}{Sokratis Katsikas}, {and} \bibinfo{person}{Catherine Meadows}} (Eds.). \bibinfo{publisher}{Springer International Publishing}, \bibinfo{address}{Cham}, \bibinfo{pages}{238--258}.
\newblock
\showISBNx{978-3-319-45744-4}


\bibitem[Yamada et~al\mbox{.}(2010)]%
        {yamada}
\bibfield{author}{\bibinfo{person}{Akira Yamada}, \bibinfo{person}{Hara Masanori}, {and} \bibinfo{person}{Yutaka Miyake}.} \bibinfo{year}{2010}\natexlab{}.
\newblock \showarticletitle{Web Tracking Site Detection Based on Temporal Link Analysis}. In \bibinfo{booktitle}{\emph{2010 IEEE 24th International Conference on Advanced Information Networking and Applications Workshops}}. \bibinfo{pages}{626--631}.
\newblock


\end{thebibliography}

\appendix

\section{Availability of Code}\label{appendix:github}
To strengthen the contribution and proper scientific conduct, we released all of our code and data online for everyone at: 

\begin{quote}
    \href{https://github.com/wolfrieder/http-response-classifier}{github.com/wolfrieder/http-response-classifier}
\end{quote}

The datasets for the year 2022~\cite{zenodo} and 2023~\cite{zenodo-new} are available under the respective references. 

\section{Evaluation Metrics}\label{appendix:metrics}
Here, we describe our additional metrics in more detail and their relevance for our classification problem. 
\begin{description}
    \item [Balanced Accuracy] Broderson et al. \cite{broderson} introduced the \textit{Balanced Accuracy} (BACC) metric for imbalanced datasets in 2010 as an alternative to average accuracy scores, which are, according to the authors, problematic for CIs and imbalanced datasets. The metric comprises \textit{recall} and \textit{true negative rate} (TNR) and is defined as: 
    \begin{equation} Balanced Accuracy = \frac{1}{2}\left(\frac{TP}{TP + FN}+\frac{TN}{TN+FP}\right)\end{equation}
    
    \item [Log-loss Score] Also referred to as cross-entropy-loss, the \textit{log-loss score} focuses on the predictive performance of a model. 
    It describes how large the error between a prediction probability and the true label is, where a value closer to zero represents a better score. However, this metric can be influenced by an imbalance, i.e.,\ the baseline log loss gets smaller with an increase in imbalance. Therefore, it is essential to compare the metric to its baseline value. For instance, the baseline for our $\text{Chrome}_{22}$ dataset with a 70:30 ratio equals 0.611~\footnote{\url{https://towardsdatascience.com/intuition-behind-log-loss-score-4e0c9979680a}}. The model's log-loss should be lower than its baseline. The log-loss score is defined as:
    \begin{equation} L_{log} (y,p) = -[y\cdot log(p) + (1-y)\cdot log(1-p)] \end{equation}
    where $y$ is the label class and $p$ is the probability.  
    
    \item [Matthews-Correlation-Coefficient] The \textit{Matthews-Correla- tion-Coefficient} (MCC)~\cite{mcc}, a metric that has seen increasing popularity in recent years, is defined as:
    \begin{equation} MCC = \frac{TP\times TN - FP \times FN}{\sqrt{(TP+FP)\cdot (TP+FN)\cdot (TN+FP)\cdot (TN+FN)}} \end{equation}
    MCC ranges from -1 to +1 and can be understood as a correlation coefficient between the true and predicted labels. The MCC will only achieve a high value if both class labels are predicted with high accuracy, positioning the MCC as a meaningful metric for imbalanced datasets \cite{mcc2}. This was further substantiated in 2020 by Chicco et al.~\cite{chicco}, who outlined how accuracy and F1-Score are not reliable measures for imbalanced datasets.  
    
    \item [AUPRC] The last metric in our study is the \textit{Area Under the Precision-Recall Curve} (AUPRC). The \textit{Precision-Recall Curve} is a plot that illustrates the trade-off between \textit{precision} and \textit{recall} for a classifier at various thresholds. The AUPRC is an aggregated measure of this curve, representing the classifier's ability to correctly identify positive instances across different thresholds. In imbalanced settings, where one class is significantly underrepresented, AUPRC is a more informative and reliable metric than ROC-AUC, as it focuses on the performance of the classifier on the minority class, which is of interest for us \cite{Saito2015ThePP}. 
\end{description}

\section{Classifier Performance without Selected Alphabet-Services}\label{appendix:robustness}
We performed an additional experiment to further test the robustness of our classifiers and to provide preliminary results for future research. The $\text{Chrome}_{22}$ classifiers were trained without HTTP responses containing \textit{Google} or \textit{DoubleClick} in their hostname and tested on the in-distribution dataset. Comparing the ET and RF with our original classifiers showed a slight decrease for AUPRC (ET: 0.959, RF: 0.962), MCC (ET: 0.876, RF: 0.872), and F1-Score (ET: 0.902, RF: 0.899). Important features were also similar, but the \textit{X-Frame-Options} header was no longer part of the top ten features and was replaced by the \textit{Priority} header. Testing the new classifiers on a dataset containing all of the previously excluded HTTP responses showed a larger decrease in performance. However, the LR model now outperformed all other classifiers -- AUPRC (ET: 0.954, RF: 0.942, LR: 0.975), MCC (ET: 0.297, RF: 0.201, LR: 0.547), and F1-Score (ET: 0.652, RF: 0.559, LR: 0.866). 

\section{Cross-Browser and HTTP Request Classifier Performance}\label{appendix:response}
This section presents our classifier performance metric results for $\text{Chrome}_{23}$, $\text{Firefox}_{22}$, $\text{Brave}_{22}$. In addition, it includes the results of our HTTP request header trained classifiers, including the top ten features across each classifier. 

\begin{table*}[t!]
	\centering
	\label{table:chrome-23-metrics}
        \tiny
	\begin{tabular}{@{}rlllllllllllll@{}}
        \toprule
		\thead{\textbf{Model}} & \thead{\textbf{Accuracy}} & \thead{\textbf{Log-Loss}} & \thead{\textbf{ROC-AUC}} & \thead{\textbf{AUPRC}} & \thead{\textbf{BACC}} & \thead{\textbf{F1-Score}} & \thead{\textbf{Precision}} & \thead{\textbf{Recall}} & \thead{\textbf{MCC}} & \thead{\textbf{FP}} & \thead{\textbf{TN}} & \thead{\textbf{FN}} & \thead{\textbf{TP}} \\
		\midrule
            LR & 0.895&0.246&0.849&0.881&0.849&0.787&0.821&0.756&0.718&34148&561918&50425&156322\\
            & [0.894;0.895]&[0.245;0.247]&[0.848;0.851]&[0.88;0.882]&[0.848;0.851]&[0.785;0.789]&[0.819;0.822]&[0.754;0.759]&[0.716;0.72] & - & - & - & - \\
            GNB & 0.86&2.47&0.825&0.759&0.825&0.735&0.717&0.754&0.64&61574&534492&50925&155822\\
            & [0.859;0.861]&[2.454;2.488]&[0.824;0.826]&[0.757;0.76]&[0.824;0.826]&[0.733;0.736]&[0.715;0.719]&[0.752;0.756]&[0.638;0.642] & - & - & - & - \\
            DT &0.928&1.358&0.905&0.854&0.905&0.86&0.861&0.859&0.811&28600&567466&29241&177506\\
            & [0.927;0.929]&[1.344;1.374]&[0.904;0.906]&[0.852;0.855]&[0.904;0.906]&[0.859;0.861]&[0.86;0.863]&[0.857;0.86]&[0.81;0.813] & - & - & - & - \\
            RF &0.936&0.326&0.913&0.94&0.913&0.875&0.883&0.866&0.832&23712&572354&27639&179108\\
            & [0.935;0.937]&[0.32;0.333]&[0.912;0.914]&[0.94;0.941]&[0.912;0.914]&[0.874;0.876]&[0.882;0.885]&[0.865;0.868]&[0.83;0.833] & - & - & - & - \\
            ET &0.944&0.381&0.92&0.936&0.92&0.888&0.905&0.873&0.851&18998&577068&26330&180417\\
            & [0.943;0.944]&[0.374;0.387]&[0.92;0.921]&[0.935;0.937]&[0.92;0.921]&[0.887;0.889]&[0.904;0.906]&[0.871;0.874]&[0.849;0.852] & - & - & - & - \\
            AdaBoost &0.891&0.565&0.848&0.873&0.848&0.782&0.809&0.757&0.711&36991&559075&50140&156607\\
            & [0.891;0.892]&[0.565;0.565]&[0.847;0.849]&[0.872;0.875]&[0.847;0.849]&[0.781;0.784]&[0.807;0.811]&[0.756;0.76]&[0.709;0.713] & - & - & - & - \\
            GBM &0.899&0.235&0.864&0.899&0.864&0.802&0.812&0.792&0.735&37790&558276&43056&163691\\
            & [0.899;0.9]&[0.234;0.236]&[0.863;0.865]&[0.898;0.9]&[0.863;0.865]&[0.801;0.803]&[0.811;0.814]&[0.789;0.794]&[0.733;0.736] & - & - & - & - \\
            LGBM &0.927&0.177&0.899&0.939&0.899&0.856&0.871&0.841&0.807&25678&570388&32912&173835\\
            & [0.926;0.928]&[0.177;0.178]&[0.898;0.9]&[0.938;0.94]&[0.898;0.9]&[0.855;0.857]&[0.87;0.873]&[0.839;0.843]&[0.806;0.809] & - & - & - & - \\
            HistGB &0.927&0.177&0.898&0.939&0.898&0.855&0.872&0.839&0.807&25504&570562&33235&173512\\
            & [0.926;0.928]&[0.176;0.178]&[0.897;0.899]&[0.938;0.939]&[0.897;0.899]&[0.854;0.857]&[0.87;0.873]&[0.837;0.841]&[0.805;0.808] & - & - & - & - \\
            XGBoost & 0.933&0.165&0.908&0.946&0.908&0.868&0.879&0.857&0.823&24323&571743&29621&177126\\
            & [0.932;0.933]&[0.164;0.166]&[0.907;0.909]&[0.946;0.947]&[0.907;0.909]&[0.866;0.869]&[0.878;0.881]&[0.854;0.858]&[0.821;0.824] & - & - & - & - \\ 
		\bottomrule
	\end{tabular}
	\\[6pt]	
 \caption{Performance comparison of the $\text{Chrome}_{\textbf{22}}$-trained classifiers that were tested on $\text{Chrome}_{\textbf{23}}$ test data.}
\end{table*}

\begin{table*}[ht]
	\centering
	\label{table:firefox-metrics}
        \tiny
	\begin{tabular}{@{}rlllllllllllll@{}}
        \toprule
		\thead{\textbf{Model}} & \thead{\textbf{Accuracy}} & \thead{\textbf{Log-Loss}} & \thead{\textbf{ROC-AUC}} & \thead{\textbf{AUPRC}} & \thead{\textbf{BACC}} & \thead{\textbf{F1-Score}} & \thead{\textbf{Precision}} & \thead{\textbf{Recall}} & \thead{\textbf{MCC}} & \thead{\textbf{FP}} & \thead{\textbf{TN}} & \thead{\textbf{FN}} & \thead{\textbf{TP}} \\
		\midrule
            LR & 0.798&0.479&0.751&0.818&0.751&0.676&0.831&0.57&0.557&35881&490984&132978&176413\\ 
            & [0.797;0.799]&[0.477;0.48]&[0.75;0.752]&[0.816;0.819]&[0.75;0.752]&[0.675;0.678]&[0.829;0.832]&[0.568;0.572]&[0.555;0.559] & - & - & - & - \\ 
            GNB & 0.789&4.382&0.754&0.703&0.754&0.685&0.765&0.62&0.535&59045&467820&117571&191820\\
            & [0.788;0.79]&[4.359;4.401]&[0.753;0.755]&[0.701;0.705]&[0.753;0.755]&[0.683;0.686]&[0.763;0.767]&[0.618;0.622]&[0.533;0.537] & - & - & - & - \\ 
            DT & 0.861&3.225&0.826&0.814&0.826&0.786&0.907&0.694&0.699&21890&504975&94749&214642\\ 
            & [0.86;0.861]&[3.207;3.248]&[0.825;0.827]&[0.813;0.816]&[0.825;0.827]&[0.785;0.788]&[0.906;0.908]&[0.692;0.695]&[0.698;0.701] & - & - & - & - \\ 
            RF & 0.854&2.835&0.822&0.892&0.822&0.78&0.884&0.698&0.684&28256&498609&93560&215831\\ 
            & [0.854;0.855]&[2.818;2.857]&[0.821;0.823]&[0.828;0.83]&[0.821;0.823]&[0.779;0.781]&[0.883;0.886]&[0.696;0.699]&[0.682;0.686] & - & - & - & - \\ 
            ET & 0.855&2.954&0.822&0.825&0.822&0.78&0.887&0.696&0.685&27377&499488&94056&21335\\ 
            & [0.854;0.856]&[2.936;2.975]&[0.821;0.823]&[0.823;0.826]&[0.821;0.823]&[0.779;0.781]&[0.886;0.888]&[0.694;0.698]&[0.683;0.687] & - & - & - & - \\ 
            AdaBoost & 0.78&0.676&0.732&0.814&0.732&0.648&0.795&0.546&0.514&43497&483368&140402&168989\\ 
            & [0.779;0.781]&[0.676;0.676]&[0.731;0.733]&[0.813;0.815]&[0.731;0.7333]&[0.646;0.649]&[0.793;0.797]&[0.544;0.548]&[0.512;0.516] & - & - & - & - \\ 
            GBM & 0.815&0.447&0.77&0.833&0.77&0.704&0.859&0.597&0.596&30267&496598&124761&184630\\ 
            & [0.814;0.815]&[0.445;0.448]&[0.769;0.771]&[0.832;0.834]&[0.769;0.771]&[0.703;0.706]&[0.858;0.861]&[0.595;0.599]&[0.594;0.598] & - & - & - & - \\ 
            LGBM &0.83&0.483&0.788&0.841&0.788&0.731&0.882&0.625&0.632&25973&500892&116153&193238\\
            & [0.829;0.831]&[0.481;0.485]&[0.787;0.789]&[0.84;0.842]&[0.787;0.789]&[0.73;0.733]&[0.88;0.883]&[0.623;0.627]&[0.63;0.633] & - & - & - & - \\ 
            HistGB &0.83&0.478&0.788&0.843&0.788&0.731&0.881&0.625&0.631&26198&500667&115984&193407\\ 
            & [0.829;0.831]&[0.476;0.48]&[0.787;0.789]&[0.842;0.844]&[0.787;0.789]&[0.73;0.733]&[0.879;0.882]&[0.624;0.627]&[0.629;0.633] & - & - & - & - \\ 
            XGBoost & 0.837&0.549&0.799&0.841&0.799&0.747&0.875&0.652&0.645&28926&497939&107618&193407\\ 
            & [0.836;0.838]&[0.547;0.552]&[0.798;0.8]&[0.839;0.841]&[0.798;0.8]&[0.746;0.749]&[0.873;0.876]&[0.651;0.654]&[0.643;0.647] & - & - & - & - \\ 
		\bottomrule
	\end{tabular}
	\\[6pt]	
 \caption{Performance comparison of the $\text{Chrome}_{\textbf{22}}$-trained classifiers that were tested on $\text{Firefox}_{\textbf{22}}$ test data.}
\end{table*}

\begin{table*}[ht]
	\centering
	\label{table:brave-metrics}
        \tiny
	\begin{tabular}{@{}rlllllllllllll@{}}
        \toprule
		\thead{\textbf{Model}} & \thead{\textbf{Accuracy}} & \thead{\textbf{Log-Loss}} & \thead{\textbf{ROC-AUC}} & \thead{\textbf{AUPRC}} & \thead{\textbf{BACC}} & \thead{\textbf{F1-Score}} & \thead{\textbf{Precision}} & \thead{\textbf{Recall}} & \thead{\textbf{MCC}} & \thead{\textbf{FP}} & \thead{\textbf{TN}} & \thead{\textbf{FN}} & \thead{\textbf{TP}} \\
		\midrule
            LR & 0.947&0.174&0.728&0.06&0.728&0.091&0.05&0.506&0.147&28500&533194&1467&1502\\ 
            & [0.946;0.948]&[0.173;0.175]&[0.719;0.738]&[0.056;0.065]&[0.719;0.738]&[0.087;0.095]&[0.048;0.052]&[0.489;0.525]&[0.141;0.153] & - & - & - & - \\ 
            GNB & 0.896&1.847&0.825&0.064&0.825&0.071&0.037&0.754&0.152&58048&503646&731&2238\\ 
            & [0.895;0.897]&[1.832;1.865]&[0.816;0.832]&[0.06;0.067]&[0.816;0.832]&[0.068;0.073]&[0.035;0.038]&[0.736;0.767]&[0.147;0.156] & - & - & - & - \\ 
            DT & 0.984&0.159&0.922&0.443&0.992&0.364&0.231&0.859&0.441&8474&553220&419&2550\\ 
            & [0.984;0.985]&[0.154;0.164]&[0.916;0.927]&[0.429;0.457]&[0.916;0.927]&[0.354;0.372]&[0.223;0.237]&[0.846;0.869]&[0.432;0.447] & - & - & - & - \\ 
            RF & 0.985&0.063&0.923&0.728&0.923&0.383&0.247&0.86&0.456&7801&552893&415&2554\\ 
            & [0.985;0.986]&[0.061;0.064]&[0.917;0.928]&[0.713;0.745]&[0.917;0.928]&[0.373;0.391]&[0.238;0.253]&[0.848;0.87]&[0.447;0.463] & - & - & - & - \\ 
            ET & 0.986&0.075&0.922&0.672&0.922&0.399&0.26&0.858&0.468&7264&554430&422&2547\\ 
            & [0.986;0.087]&[0.073;0.078]&[0.916;0.927]&[0.654;0.687]&[0.916;0.927]&[0.387;0.407]&[0.251;0.266]&[0.846;0.866]&[0.458;0.474] & - & - & - & - \\ 
            AdaBoost & 0.94&0.669&0.734&0.042&0.734&0.085&0.046&0.525&0.142&32299&529395&1411&1558\\ 
            & [0.94;0.941]&[0.669;0.669]&[0.725;0.742]&[0.039;0.045]&[0.725;0.742]&[0.081;0.088]&[0.044;0.048]&[0.508;0.542]&[0.137;0.148] & - & - & - & - \\ 
            GBM & 0.963&0.158&0.835&0.439&0.835&0.167&0.095&0.706&0.25&19964&541730&873&2096\\ 
            & [0.963;0.964]&[0.157;0.158]&[0.829;0.843]&[0.418;0.46]&[0.829;0.843]&[0.162;0.172]&[0.092;0.098]&[0.693;0.72]&[0.244;0.256] & - & - & - & - \\ 
            LGBM & 0.979&0.109&0.86&0.537&0.86&0.267&0.163&0.74&0.341&11312&550382&773&2196\\ 
            & [0.978;0.979]&[0.109;0.11]&[0.851;0.867]&[0.518;0.559]&[0.851;0.867]&[0.257;0.274]&[0.155;0.168]&[0.724;0.755]&[0.33;0.349] & - & - & - & - \\ 
            HistGB & 0.979&0.109&0.861&0.549&0.861&0.267&0.163&0.742&0.341&11340&550354&766&2203\\ 
            & [0.978;0.979]&[0.108;0.109]&[0.854;0.869]&[0.531;0.57]&[0.854;0.869]&[0.256;0.275]&[0.156;0.168]&[0.727;0.758]&[0.331;0.349] & - & - & - & - \\ 
            XGBoost & 0.979&0.091&0.871&0.643&0.871&0.276&0.169&0.762&0.352&11148&550546&707&2262\\ 
            & [0.979;0.979]&[0.091;0.092]&[0.864;0.878]&[0.627;0.661]&[0.864;0.878]&[0.267;0.284]&[0.162;0.174]&[0.747;0.775]&[0.343;0.361] & - & - & - & - \\ 
		\bottomrule
	\end{tabular}
	\\[6pt]	
 \caption{Performance comparison of the $\text{Chrome}_{\textbf{22}}$-trained classifiers that were tested on $\text{Brave}_{\textbf{22}}$ test data.}
\end{table*}

\begin{figure*}[htbp] 
    \centering
    \includegraphics[width=0.9\textwidth]{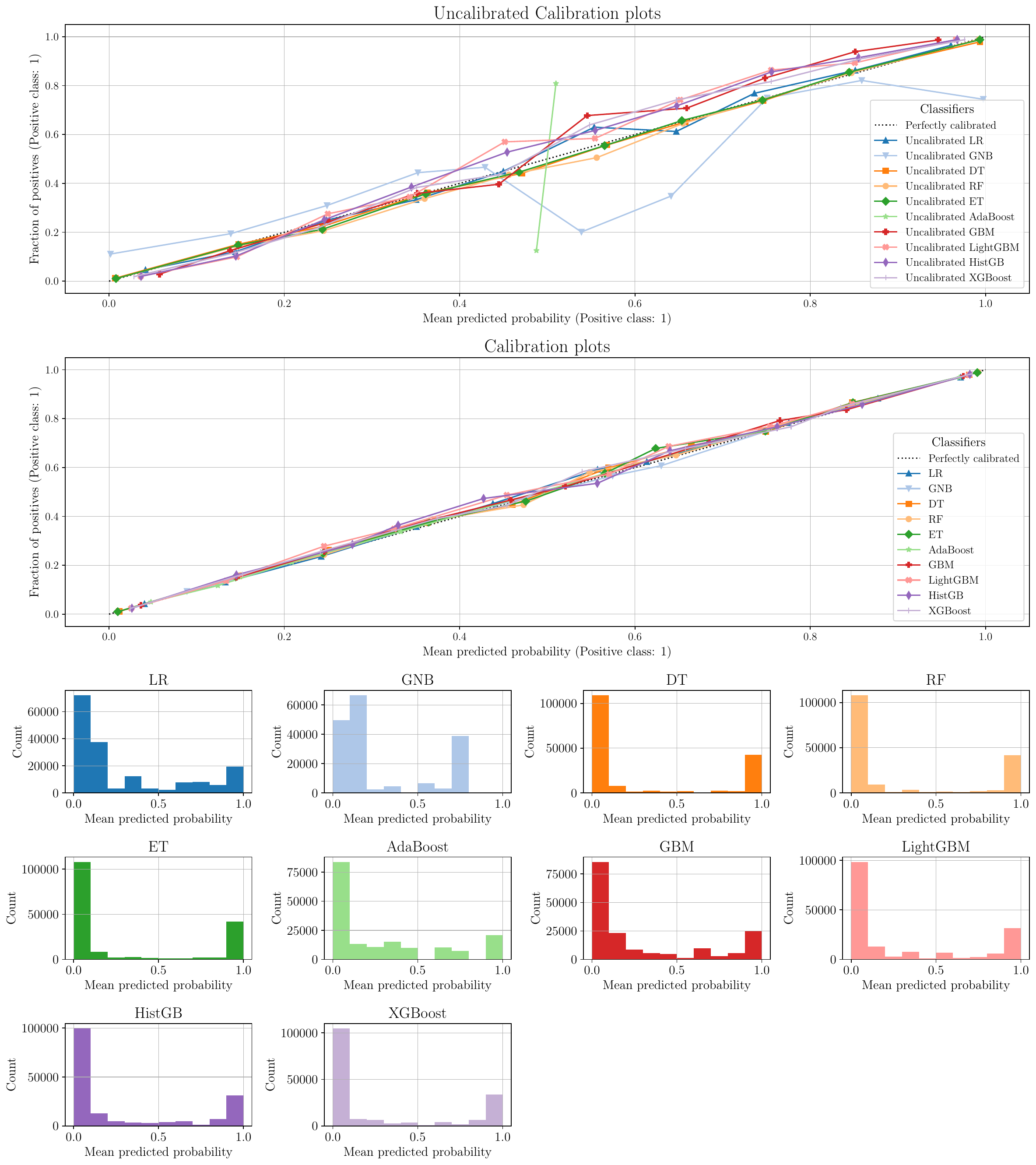}
    \caption{The AdaBoost and GNB classifier improved significantly through calibration using isotonic regression.}
    \label{fig:calib}
    \Description[Calibrating the classifiers for potential perfomance improvements.]{The AdaBoost and GNB classifier improved significantly through calibration using isotonic regression.}
\end{figure*}



\begin{figure*}[ht!] 
    \centering
    \includegraphics[width=0.8\textwidth]{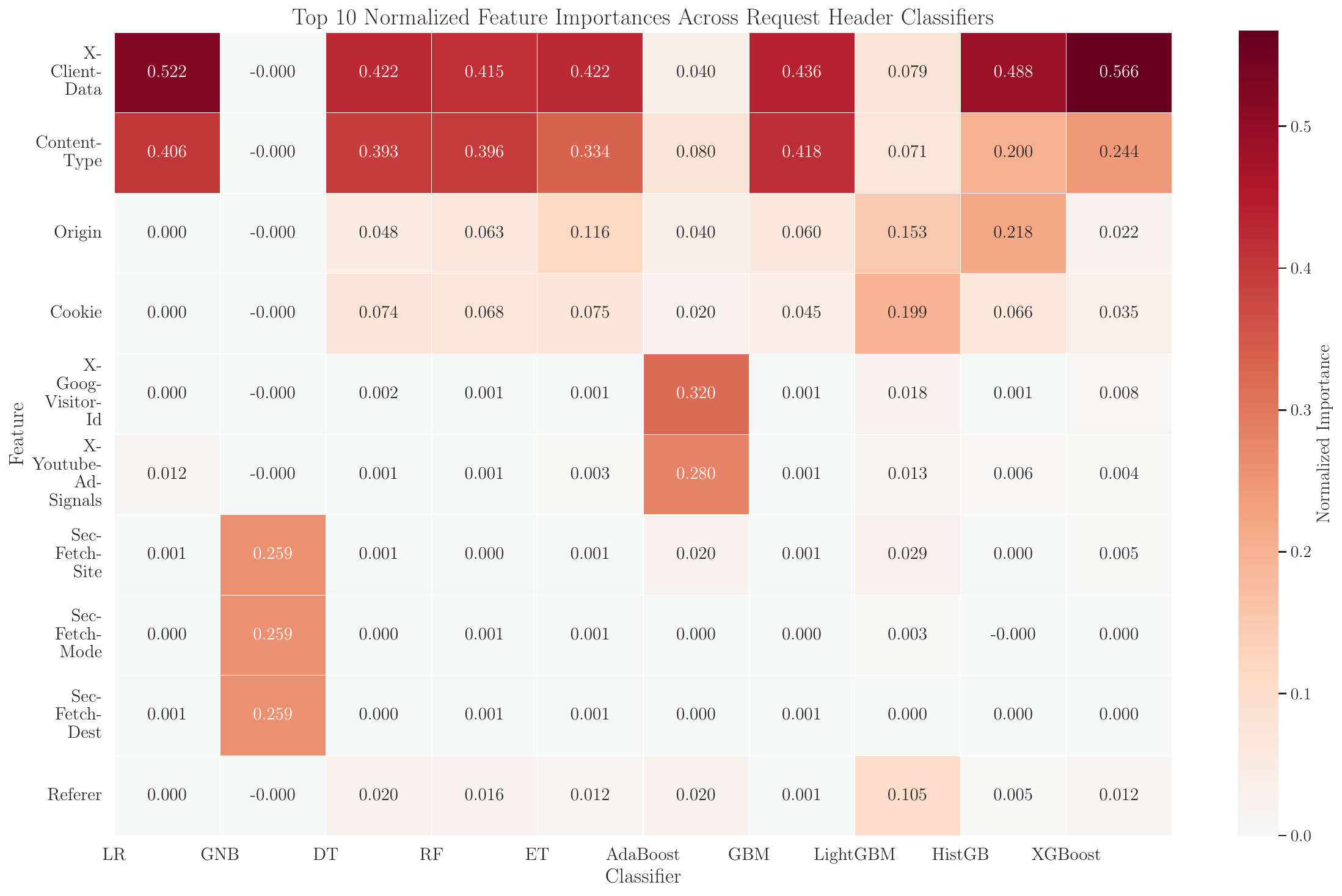}
    \caption{The top ten features are presented in order of importance across different classifiers based on the in-distribution test set. The heatmap illustrates the normalized importance values for each feature and classifier. We calculated \textit{Permutation feature importance} for LR and GNB but not for the others as we do not have any high cardinality features.} 
    \label{fig:feature_importance}
    \Description[Feature importance for the request-based classifiers.]{The top ten features are presented in order of importance across different classifiers based on the in-distribution test set. The heatmap illustrates the normalized importance values for each feature and classifier. We calculated \textit{Permutation feature importance} for LR and GNB but not for the others as we do not have any high cardinality features.}
\end{figure*}

\begin{table*}[ht]
	\centering
	\label{table:chrome-request-metrics}
        \tiny
	\begin{tabular}{@{}rlllllllllllll@{}}
        \toprule
		\thead{\textbf{Model}} & \thead{\textbf{Accuracy}} & \thead{\textbf{Log-Loss}} & \thead{\textbf{ROC-AUC}} & \thead{\textbf{AUPRC}} & \thead{\textbf{BACC}} & \thead{\textbf{F1-Score}} & \thead{\textbf{Precision}} & \thead{\textbf{Recall}} & \thead{\textbf{MCC}} & \thead{\textbf{FP}} & \thead{\textbf{TN}} & \thead{\textbf{FN}} & \thead{\textbf{TP}} \\
		\midrule
            LR & 0.767&0.537&0.663&0.515&0.663&0.509&0.687&0.404&0.392&9420.0&111084.0&30559.0&20682.0\\ 
            & [0.765;0.769]&[0.534;0.539]&[0.66;0.665]&[0.511;0.52]&[0.66;0.665]&[0.504;0.513]&[0.683;0.693]&[0.4;0.408]&[0.387;0.397] & - & - & - & - \\ 
            GNB & 0.304&3.056&0.502&0.488&0.502&0.46&0.299&0.995&0.025&119366.0&1138.0&236.0&51005.0\\ 
            & [0.302;0.305]&[3.037;3.074]&[0.502;0.503]&[0.484;0.493]&[0.502;0.503]&[0.458;0.463]&[0.298;0.301]&[0.995;0.996]&[0.021;0.029] & - & - & - & - \\ 
            DT & 0.778&0.521&0.667&0.548&0.667&0.514&0.743&0.393&0.42&6965.0&113539.0&31123.0&20118.0\\ 
            & [0.776;0.78]&[0.518;0.523]&[0.665;0.67]&[0.543;0.552]&[0.665;0.67]&[0.509;0.518]&[0.738;0.749]&[0.388;0.398]&[0.416;0.425] & - & - & - & - \\ 
            RF & 0.778&0.52&0.667&0.548&0.667&0.514&0.743&0.393&0.42&6969.0&113535.0&31113.0&20128.0\\ 
            & [0.776;0.78]&[0.518;0.523]&[0.665;0.67]&[0.543;0.552]&[0.665;0.67]&[0.509;0.519]&[0.738;0.749]&[0.388;0.398]&[0.416;0.425] & - & - & - & - \\ 
            ET & 0.778&0.52&0.667&0.548&0.667&0.514&0.743&0.393&0.42&6966.0&113538.0&31123.0&20118.0\\ 
            & [0.776;0.78]&[0.518;0.523]&[0.665;0.67]&[0.543;0.552]&[0.665;0.67]&[0.509;0.518]&[0.738;0.749]&[0.388;0.398]&[0.416;0.425] & - & - & - & - \\ 
            AdaBoost & 0.767&0.686&0.663&0.516&0.663&0.508&0.687&0.404&0.392&9423.0&111081.0&30562.0&20679.0\\ 
            & [0.765;0.769]&[0.686;0.686]&[0.66;0.665]&[0.511;0.52]&[0.66;0.665]&[0.504;0.513]&[0.682;0.692]&[0.4;0.408]&[0.387;0.397] & - & - & - & - \\ 
            GBM & 0.777&0.525&0.666&0.539&0.666&0.512&0.737&0.393&0.417&7183.0&113321.0&31124.0&20117.0\\ 
            & [0.775;0.779]&[0.523;0.528]&[0.664;0.669]&[0.535;0.543]&[0.664;0.669]&[0.507;0.517]&[0.732;0.742]&[0.388;0.397]&[0.412;0.422] & - & - & - & - \\ 
            LGBM & 0.778&0.52&0.667&0.548&0.667&0.514&0.743&0.393&0.42&6968.0&113536.0&31123.0&20118.0\\ 
            & [0.776;0.78]&[0.518;0.523]&[0.665;0.67]&[0.543;0.552]&[0.665;0.67]&[0.509;0.518]&[0.738;0.749]&[0.388;0.398]&[0.416;0.425] & - & - & - & - \\ 
            HistGB & 0.778&0.52&0.667&0.548&0.667&0.514&0.743&0.393&0.42&6971.0&113533.0&31115.0&20126.0\\ 
            & [0.776;0.78]&[0.518;0.523]&[0.665;0.67]&[0.543;0.552]&[0.665;0.67]&[0.509;0.519]&[0.738;0.749]&[0.388;0.398]&[0.416;0.425] & - & - & - & - \\ 
            XGBoost & 0.778&0.52&0.667&0.548&0.667&0.514&0.743&0.393&0.42&6966.0&113538.0&31123.0&20118.0\\ 
            & [0.776;0.78]&[0.518;0.523]&[0.665;0.67]&[0.543;0.552]&[0.665;0.67]&[0.509;0.518]&[0.738;0.749]&[0.388;0.398]&[0.416;0.425] & - & - & - & - \\ 
		\bottomrule
	\end{tabular}
	\\[6pt]	
 \caption{Performance comparison of the $\text{Chrome}_{\textbf{22}}$-trained classifiers with HTTP request headers that were tested on in-distribution test data.}
\end{table*}










\end{document}